\documentclass[onecolumn,11pt,amssymb,amsmath,preprintnumbers,nofootinbib,superscriptaddress,aps]{revtex4-1}
\usepackage{lineno}
\usepackage{bm,color}
\usepackage[dvipsnames]{xcolor}
\usepackage{slashed} 
\usepackage{graphicx}
\usepackage{soul} 
\usepackage{multirow}
\usepackage{comment}
\usepackage{mathtools}
\usepackage{appendix}
\usepackage[colorlinks=true
,urlcolor=blue
,anchorcolor=blue
,citecolor=blue 
,filecolor=blue
,linkcolor=blue
,menucolor=blue
,linktocpage=true
,pdfproducer=medialab
,pdfa=true
]{hyperref}
\usepackage{lineno}

\newcommand{\GeV}{\ensuremath{\mathrm{GeV}}}

\newcommand{\tw}{\theta_W}

\newcommand{\beq}{\begin{equation}}
\newcommand{\eeq}{\end{equation}}
\newcommand{\bea}{\begin{eqnarray}}
\newcommand{\eea}{\end{eqnarray}}

\begin{document}

 
\begin{flushright}
\end{flushright}

\title{\bf \Large Speculations on the W-Mass Measurement at CDF}



\author{Jiayin Gu}
\email{jiayin\_gu@fudan.edu.cn}
\affiliation{Department of Physics and Center for Field Theory and Particle Physics, Fudan University, Shanghai 200438, China}
\affiliation{Key Laboratory of Nuclear Physics and Ion-beam Application (MOE), Fudan University, Shanghai 200433, China}
\author{Zhen Liu}
\email{zliuphys@umn.edu}
\thanks{\scriptsize \!\! \href{https://orcid.org/0000-0002-3143-1976}{0000-0002-3143-1976}}
\affiliation{\it School of Physics and Astronomy, University of Minnesota, Minneapolis, MN 55455, USA}
\author{Teng Ma}
\email{t.ma@campus.technion.ac.il}
\affiliation{Physics Department, Technion – Israel Institute of Technology, Haifa 3200003, Israel
}
\author{Jing Shu}
\email{jshu@itp.ac.cn}
\affiliation{
CAS Key Laboratory of Theoretical Physics, Institute of Theoretical
Physics, Chinese Academy of Sciences, Beijing 100190, P.R.China}
\affiliation{School of Physical Sciences, University of Chinese Academy of Sciences, Beijing 100049, China}
\affiliation{School of Fundamental Physics and Mathematical Sciences, Hangzhou Institute for Advanced Study, University of Chinese Academy of Sciences, Hangzhou 310024, China}
\affiliation{International Center for Theoretical Physics Asia-Pacific, Beijing/Hangzhou, China
}


\begin{abstract}
The W Mass determination at the Tevatron CDF experiment reported a deviation from the SM expectation at 7$\sigma$ level. We discuss a few possible interpretations and their collider implications. We perform electroweak global fits under various frameworks and assumptions. We consider three types of electroweak global fits in the effective-field-theory framework: the $S$-$T$, the $S$-$T$-$\delta G_F$, and the eight-parameter flavor-universal one. We discuss the amounts of tensions between different $m_W$ measurements reflected in these fits and the corresponding shifts in central values of these parameters. With these electroweak fit pictures in hand, we present a few different classes of models and discuss their compatibility with these results. We find that while explaining the $m_W$ discrepancy, the single gauge boson extensions face strong LHC direct search constraints unless the $Z'$ is fermiophobic (leptophobic) which can be realized if extra vector fermions (leptons) mix with the SM fermions (leptons). Vector-like top partners can partially generate the needed shift to the electroweak observables. The compatibility with top squark is also studied in detail. We find non-degenerate top squark soft masses enhance the needed operator coefficients, enabling an allowed explanation compatible with current LHC measurements. Overall, more theory and experimental developments are highly in demand to reveal the physics behind this discrepancy.
\end{abstract}

\maketitle




{\small 
\tableofcontents}

\section{Introduction}

The $W$ mass measurement recently published by the CDF~II experiment~\cite{CDF:2022hxs} reports 7 sigma deviation from its standard model (SM) expectation (obtained from an electroweak global fit), and is also in tension with other experimental determinations~\cite{Zyla:2020zbs}. The new CDF-II result, the SM expectation from the electroweak fit, and the previous PDG world-average value are~\cite{CDF:2022hxs, Zyla:2020zbs}
\begin{eqnarray}
m_W^{\rm CDF-II}&=&80433.5\pm 9.4~{\rm MeV}\,, \nonumber\\
m_W^{\rm EW-fit}&=&80357\pm 6~{\rm MeV}\,, \nonumber \\
m_W^{\rm PDG}&=& 80379 \pm 12~{\rm MeV}  \,. \nonumber 
\end{eqnarray}
The standard model of particle physics is a precise and concise theory. The fact that we are testing it from many different angles enables us to understand where new physics could come. 
This new CDF~II $W$ mass measurement, being the most precise one so far,  
is certainly a remarkable achievement, and the result itself calls for a lot of new explorations in particle physics. In this paper, we study and discuss various possibilities of the physics behind such discrepancies in the measurements.

Notably, this new determination of $m_W$ took advantage of the large integrated luminosity, the kinematics of proton-antiproton collision, and the high energy resolution level without an exceedingly large pile-up. Through a template fit to the underlying $W$ boson mass, the measurement is robustly checked against various experimental effects. The result is impressively precise. Several outstanding theories and experimental directions can make the results and interpretation more robust. This measurement is in tension with other experimental measurements from different collaborations~\cite{ALEPH:2010aa,CDF:2013dpa,ATLAS:2017rzl,Zyla:2020zbs,LHCb:2021bjt},  
which calls for comparative studies.  
An understanding of various experimental assumptions and calibrations is of crucial importance. In particular, the $W$ boson production and decays are subject to various high order QCD and QED effects, both fixed order and resumed. The differential rate itself has sizable scale uncertainties, resulting from the missing higher-order calculations. It would be useful to understand the theoretical uncertainties behind these ``templates" 
used by the CDF-II experiment. Only with a sufficiently precise theoretical control can the measured $W$-boson mass be interpreted as a well-defined quantity to be compared with the predictions from the SM and new physics.  

\begin{table}[h]
  \centering
    \begin{tabular}{c|c|c|c|c|c|c}
    \multicolumn{2}{c|}{\multirow{2}[0]{*}{Specifications/models}} & \multirow{2}[0]{*}{d.o.f.} & \multicolumn{3}{c|}{$\chi^2$} & \multirow{2}[0]{*}{References} \\ \cline{4-6}
    \multicolumn{2}{c|}{} &   & pre CDF-II & $m_W^{\rm combine}$ & $m_W^{\rm CDF-II}$ &  \\ \hline\hline
    \multirow{4}[0]{*}{EW fit} & SM & (3) & 31 & 62  & 76 & \multirow{4}[0]{*}{\autoref{sec:EWfit}} \\ \cline{2-6}
      & S-T & (3)+2 &  28 & 30  & 33  &  \\ \cline{2-6}
      & S-T-$\delta G_F$ & (3)+3 & 28 & 28  & 28  &  \\ \cline{2-6}
      & Universal EW & (3)+8 & 17  & 17  & 17  &  \\ \cline{2-6}
      \hline\hline
    \multirow{3}[0]{*}{BSM Models}  
      & $Z^\prime$/$W^\prime$ ($\Delta S=0.1$)\footnote{$\Delta S=0.1$ in the bracket represents the case that extra contribution to oblique parameter $S$ is included, and the corresponding best fit $\chi^2$s are listed in the parathensis.} & (3)+1\footnote{While the model has more free parameters, with the gauge symmetry breaking assumption, only one linear combinations of parameters enters the fit.}& 29 (28)  & 38 (33) & 34 (31) &  \autoref{sec:zprime} \\ \cline{2-7}
     & VLQ Top I ($\Delta S=0.1$) & (3)+2\footnote{In the simplified singlet and doublet top partner models, there are only two free parameters, $M_{T}$ ($M_{Q_T}$) and $\lambda_{T}$ ($\lambda_{Q_T}$).} & 29 (29) & 34 (32) &   38 (34) &  \autoref{sec:toppartner1} \\ \cline{2-7}
      & VLQ Top II ($\Delta S=0.1$) & (3)+2 & 28 (53)  & 33 (31)  & 37 (33)  &  \autoref{sec:toppartner2} \\ \cline{2-7}
      & Top Squark & (3)+2\footnote{In the degenerate soft-mass scenario, only two degree of freedom of top squarks, $m_{\tilde t}$ and $\tan\beta$ are present. In the non-degenerate soft-mass scenarios, three degrees of freedom, $m_{\tilde Q_3}$, $m_{\tilde U_3}$, $\tan\beta$ are present. In both scenarios, $\tan\beta$ does not change the quality of the fit in a sizable way.} & 28  & 31  & 34  &  \autoref{sec:stop} \\ \hline
    \end{tabular}%
    \caption{The summary of various SM and BSM considerations in this study with the corresponding references. The best fit $\chi^2$s are listed for various scenarios considered here in this study. For BSM considerations, the best fit values of individual models are shown. However, other direct experimental searches need to be taken into account and these details are discussed in the text. 
    }
  \label{tab:summary}%
\end{table}%

Cautioned by the further experimental and theoretical work needed to fully establish this discrepancy in $W$ mass determination, these intriguing results call for evaluations of possible new physics sources. Should it come from new physics, plausible BSM scenarios and testable aspects are presented in this work. We discuss the overall picture of Electroweak Precision Observables~(EWPO) fit and how several simple, representative models would be able to help improve the fitting. A brief summary is outlined below. 
Ultimately, a new global average of the $W$-mass measurements that includes the new CDF-II result should be obtained and used in the global analyses.  The combination of $W$-mass measurements is however highly nontrivial given the sizable amount of tension between the CDF-II measurement and the previous ones.  In our study, we mainly focus on the comparison between the previous world-average $m_W$ measurement and the new CDF-II one rather than their combination.  
%

This paper is organized as the following. In \autoref{sec:EWfit}, we perform three types of electroweak (EW) global fits in an effective-field-theory framework, namely the $S$-$T$, the $S$-$T$-$\delta G_F$, and the eight-parameter flavor-universal one. Each of them is distinct in terms of correlations, best-fit values, and the amount of tensions between EW and this new $m_W$ determination. In \autoref{sec:BSM} we introduce three classes of models, each accompanied with two scenarios, that we check if they could help reduce the tension between SM EW fit and this new direct $m_W$ measurement. The models we studied include gauge extensions of the SM with new gauge bosons, vector-like top partners, and top squarks. These results are summarized in~\autoref{tab:summary}. We further check these intuitive models' theory and experimental constraints and comment on their future perspectives. Finally in \autoref{sec:outlook}, we conclude.

\section{EW Fit}
\label{sec:EWfit}


Here we provide the interpretation of a shifted $m_W$ from an effective-field-theory point of view.  We work in the framework of the Standard Model Effective Field Theory~\cite{Buchmuller:1985jz, Grzadkowski:2010es}. 
We choose the $\{ \alpha,\, m_{Z}, G_{F}\}$ input scheme so that the measurement of $m_W$ provides a constraint on the operator coefficients.  More specifically, we fix the measured values of these input parameters to be~\cite{Zyla:2020zbs}, 
\begin{equation}
\alpha=1/127.940,\,~~~~ m_{Z}=91.1876\,\GeV, \,~~~~ G_{F}=1.1663787\times 10^{-5}\,\GeV \,.
\end{equation}
Any new physics effects that contributes to the measurement of these parameters thus change the ``inferred SM value'', and contribute indirectly to the observables.  For instance, the 4-fermion operator $\mathcal{O}^{1221}_{\ell\ell} =  (\bar{\ell}_1 \gamma^\mu \ell_2) (\bar{\ell_2} \gamma_\mu \ell_1)$ contributes to the muon decay process, and generates a shift in the inferred SM $G_F$ (VEV\footnote{Vacuum Expectation Value}) which we will later denote as $\delta G_F$.  
We focus on the Z and W pole measurements in our analysis, which are 
\begin{equation}
\Gamma_Z \,, \hspace{0.5cm}
\sigma_{\rm had} \,, \hspace{0.5cm}
R_{f} \,,  \hspace{0.5cm}
A^{0,\, f}_{\rm FB} \,, \hspace{0.5cm}
A_{f}  \,, \hspace{0.5cm}
A^{\rm pol}_{e/\tau}  \,, 
\end{equation}
where $f=e,u,\tau,b,c$, and $A^{\rm pol}_{e/\tau}$  are $A_e$ and $A_\tau$ measured from tau polarization measurements at LEP, and
\begin{equation}
m_W \,, \hspace{0.5cm}
\Gamma_W  \,, \hspace{0.5cm}
{\rm BR}(W\to e\nu)  \,, \hspace{0.5cm}
{\rm BR}(W\to \mu\nu)  \,, \hspace{0.5cm}
{\rm BR}(W\to \tau\nu) \,.
\end{equation}
For the Z-pole measurements, we use the results in Ref.~\cite{ALEPH:2005ab}, where the correlations (if available) are also included.  The measurements of W branching ratios are taken from Ref.~\cite{ALEPH:2013dgf}.  For $\Gamma_W$, we use the PDG result~\cite{Zyla:2020zbs}. 
For $m_W$, 
we consider two scenarios, the ``old'' world average from PDG~\cite{Zyla:2020zbs}, and the new CDF measurement~\cite{CDF:2022hxs} alone. A ``new" combination with between other experiments and the CDF results are not considered here, although it can be done straightforwardly, as our purpose is to evaluate the new physics needed if CDF result is the correct value of $m_W$. 

The SMEFT parameterization in our analysis follow closely the ones in Refs.~\cite{Falkowski:2014tna, Efrati:2015eaa}~(see also Refs.~\cite{Ellis:2018gqa, Dawson:2020oco, Ellis:2020unq, Ethier:2021bye, Almeida:2021asy}), where the contributions to observables from the dimension-6 operators are calculated at the tree-level, but normalized to the SM predictions.  The SM predictions are taken from the central values of the SM-fit in Ref.~\cite{Baak:2014ora}, except for $m_W$, which is from~Ref.~\cite{Zyla:2020zbs}.  To account for parametric and theory uncertainties that are absent in this simple treatment, we combine in quadrature the experimental uncertainty of $m_W$ with the one from the ``SM EW fits''~\cite{Zyla:2020zbs}
, treating the latter as an effective total ``theory'' error.  
This theory error, 6\,MeV, mainly comes from the missing higher-order calculations and the parametric uncertainties of input parameters, which include $m_t$ and $m_H$ that enter at the one-loop level.  
Our results from this simple treatment for the $S,\,T$ parameters are in good agreement with the ones from Ref.~\cite{Baak:2014ora}. 
%


Before doing a detailed analysis, it is intuitive to first try to understand what kind of new physics contribution could generate a significant shift on $m_W$ without modifying any other electroweak observable, as the latter is generally in good agreement with the SM predictions~\cite{ALEPH:2005ab}. It is convenient to work in a basis where the operators associated with the $W$, $Y$ parameters~\cite{Barbieri:2004qk} are exchanged for the 4-fermion operators.  In this case, the modification of $m_W$ from dimension-6 operators is given by
%
\begin{equation}
\label{eq:fits}
    \delta m_W = \frac{1}{2 c_{2w}} \left[ c^2_w \, \hat{T} - s^2_w \left( \delta G_F + 2 \hat{S} \right) \right] \,,
\end{equation}
where $m_W = m_W^{\rm SM}(1+\delta m_W)$, $G_F = G_F^{\rm SM}(1+\delta G_F)$,\footnote{Here $m_W$ and $G_F$ are the measured values, $G_F^{\rm SM}$ is the inferred SM value and $m_W^{\rm SM}$ is the SM prediction.} $s^2_w \equiv \sin^2\theta_W$, $c^2_w \equiv \cos^2\theta_W$, $c_{2w} \equiv \cos2\theta_W$, and $\theta_W$ is the weak-mixing angle.  The parameters $\hat{S}$ and $\hat{T}$~\cite{Barbieri:2004qk} are related to the $S$ and $T$ parameters~\cite{Peskin:1991sw} by
\begin{equation}
\hat{S} = \frac{\alpha}{4 s^2_w} S \,, \hspace{1.5cm}
\hat{T} = \alpha T \,.
\end{equation}
Note also that the $U$ parameter is generated by dimension-8 operators and is not considered here.  Among the three parameters $\hat{S}$, $\hat{T}$ and $\delta G_F$, only two independent combinations contribute to the Z-pole observables, which are $\hat{T}-\delta G_F$ and $\hat{S}$, respectively.  Therefore, to generate a positive $\delta m_W$ without changing the Z-pole observables, one needs to keep $\hat{S}=0$, and shift $\hat{T}$ and $\delta G_F$ simutenously such that $\hat{T} = \delta G_F$.  A positive $\hat{T}$ (and $\delta G_F$) is needed for a positive $\delta m_W$.

\begin{figure}
    \centering
    \includegraphics[width=0.48\textwidth]{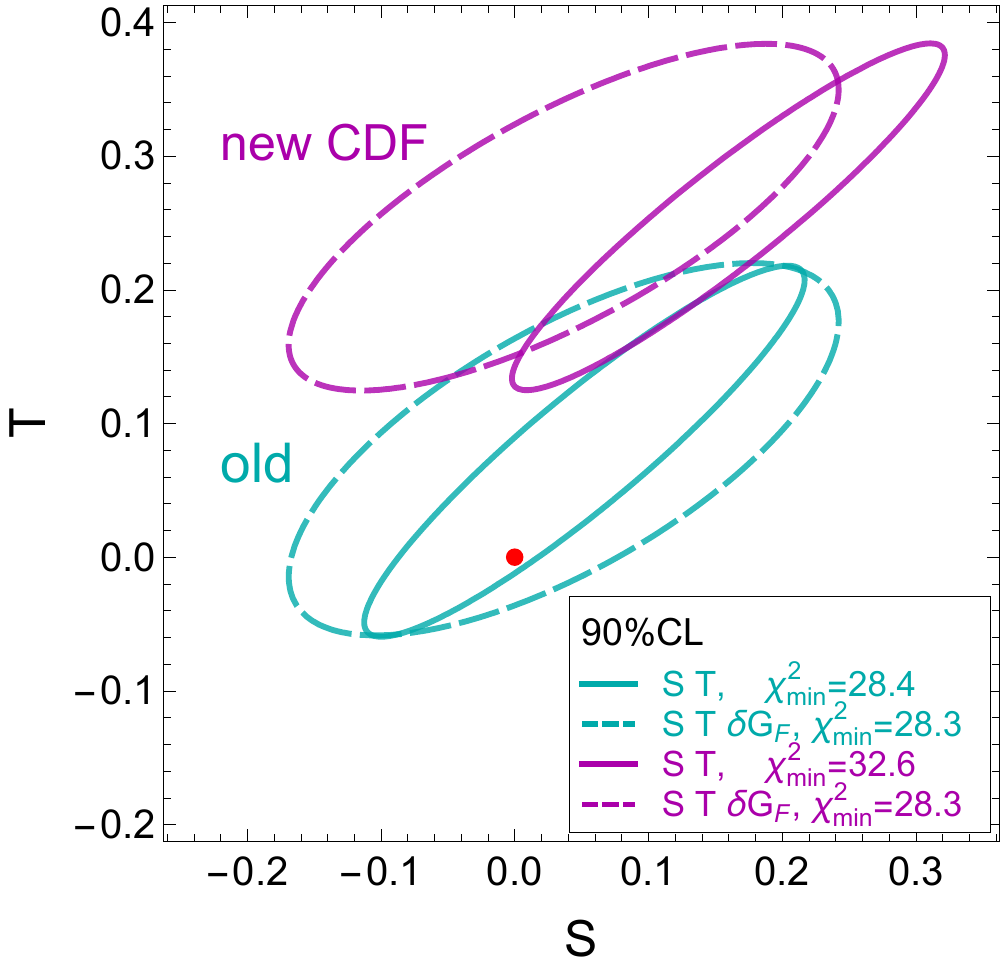}
    \includegraphics[width=0.48\textwidth]{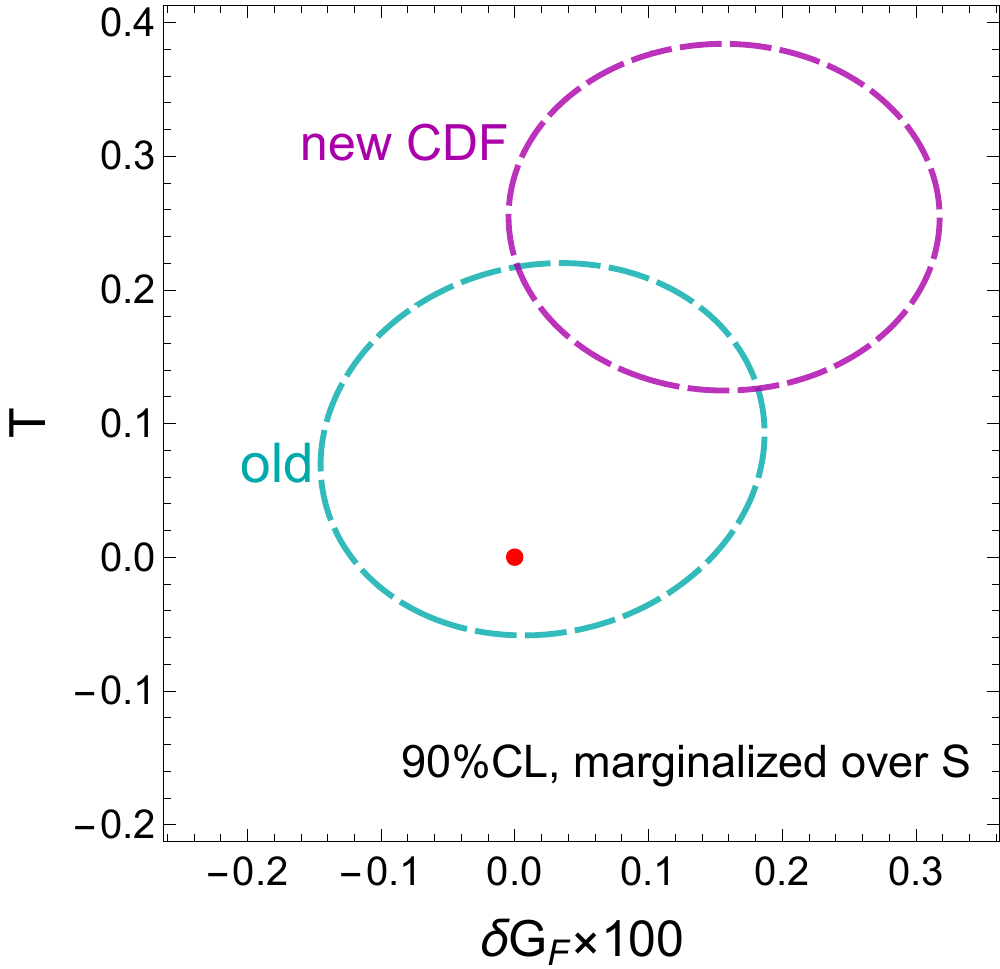}
    \caption{Results from a 2-parameter fit of $S$ and $T$ (solid contours) and a 3-parameter fit of $S$, $T$ and $\delta G_F$ (dashed contours) to the current EW precision measurements.  The ``old'' scenario (cyan) uses the current PDG world-average $m_W$ measurement, while the ``new CDF'' scenario (magenta) uses the new CDF measurement alone for $m_W$.  We fix $U=0$ as it is generated by dimension-8 operators.  Left (right) panel shows the results in the $S$-$T$ ($\delta G_F$-$T$) plane.  The scale for $\delta G_F$ is amplified by 100 for convenience.  All contours correspond to 90\%\,CL.
    }
    \label{fig:STfit}
\end{figure}

To verify this statement, we first perform a global fit of the 3 parameters above. 
The results are presented in \autoref{fig:STfit}.  
The ``old'' scenario, with the world-average $m_W$ measurement, is compared with the ``new CDF'' scenario, with $m_W$ from only the new CDF~II measurement.  For easy comparison, we switch to the original (no hat) version of $S$ and $T$.  The left panel shows the 90\% confidence level (CL) contour in the $S$-$T$ plane.  For each scenario, two contours are shown: the solid one is from a 2-parameter fit of $S$ and $T$, setting $\delta G_F =0$, while the dashed one is from the 3-parameter fit (marginalized over $G_F$).  For the 3-parameter fits, the contours are also projected on the $(\delta G_F,\,T)$ plane shown on the right panel. The results are also listed in \autoref{tab:STG}.

From \autoref{fig:STfit} we could see that, indeed, the shift of the central values in the $(S,T,\delta G_F)$ 3-parameter fit is consistent with our observation above.  From the old $m_W$ to the new CDF one, the central values of $T$ and $\delta G_F$ both shift in the positive direction, while the central value of $S$ does not change.  We also note that the minimum $\chi^2$ are the same for the old and new scenarios in this case.  On the other hand, for the $(S,T)$ 2-parameter fit, shifts of both $S$ and $T$ in the positive direction are required to go from the old $m_W$ measurement to the new CDF one.  The minimum $\chi^2$ is also increased by around 4, suggesting that the $S,T$-only scenario exhibits some tension with the new CDF measurement. 

\begin{table}
\small
\centering
\begin{tabular}{|c||c|ccc||c|ccc|} \hline
 &\multicolumn{4}{|c||}{old (PDG)} & \multicolumn{4}{|c|}{new CDF}  \\  \hline\hline
2-para & $1\sigma$ bound &   \multicolumn{3}{c||}{correlation matrix}    & $1\sigma$ bound &   \multicolumn{3}{c|}{correlation matrix}    \\ \cline{2-9}
fit &  &    $~S~$  &  $~T~$  &   &  &   $~S~$  &  $~T~$  &      \\ \hline
$S$    &    $0.052 \pm 0.077$    &    1 & 0.92 &  &   $0.160\pm 0.075$    &    1 & 0.93 &    \\
$T$     &    $0.079 \pm 0.065$    &       &  1      &    &    $0.255 \pm 0.060$    &       &  1      &      \\ \hline\hline
3-para & $1\sigma$ bound &   \multicolumn{3}{c||}{correlation matrix}    & $1\sigma$ bound &   \multicolumn{3}{c|}{correlation matrix}    \\ \cline{2-9}
fit &  &    $~S~$  &  $~T~$  & $\delta G_F$ &  &   $~S~$  &  $~T~$  & $\delta G_F$     \\ \hline
$S$    &    $0.037 \pm 0.096$    &    1 & 0.68 & -0.60  &   $0.037\pm 0.096$    &    1 & 0.74 & -0.62   \\
$T$     &    $0.081 \pm 0.065$    &       &  1      &  0.08  &    $0.254 \pm 0.060$    &       &  1      &  -0.002     \\
$\delta G_F$     &      $(2.1 \pm 7.7)\times 10^{-4}$   &       &          & 1 &     $(15.7 \pm 7.5)\times 10^{-4}$   &       &          & 1         \\  \hline
\end{tabular}
\caption{Results from the $(S,T)$ 2-parameter fit and $(S,T,\delta G_F)$ 3-parameter fit as in \autoref{fig:STfit}.  One-sigma bounds are quoted here.}
\label{tab:STG}
\end{table}

Let us now move to a more general framework with a complete basis of 8 operators for the $Z$ and $W$ pole observables, assuming flavor universality.  
The corresponding Lagrangian is given by
\begin{equation}
    \mathcal{L} = \frac{c_{WB}}{ m^2_W} \mathcal{O}_{WB}  + \frac{c_T}{v^2} \mathcal{O}_{T}  +  \frac{c^{1221}_{ll}}{v^2} \mathcal{O}^{1221}_{\ell\ell} + \frac{c'_{Hq}}{v^2} \mathcal{O}'_{Hq} + \underset{f=e,q,u,d}{\sum} \frac{c_{Hf}}{v^2} \mathcal{O}_{Hf} \,, \label{eq:Lsmeft}
\end{equation}
where the operators are listed in \autoref{tab:op1}, and $v\simeq 246$\,GeV.  Note that we keep the flavor labels on the 4-fermion operator $\mathcal{O}^{1221}_{\ell\ell}$.  This is because the flavor diagonal ones (with indices $iijj$), which can for instance be generated by flavor-preserving interactions with a $Z'$-boson, do not contribute to the muon decay (and $\delta G_F$).  This information is somewhat unclear under the flavor universality condition, as the $\mathcal{O}^{ijji}_{\ell\ell}$ operator can be expressed as a combination of flavor diagonal operators using the Fiertz identity. 
We also choose the convention for $\mathcal{O}_{WB}$ to have $c_{WB} = \hat{S}$ (which is different from the convention in {\it e.g.} Ref.~\cite{DeBlas:2019qco}).  This basis matches the SILH' basis in Refs.~\cite{Elias-Miro:2013mua, Pomarol:2013zra} with the replacement $c_{WB} \to c_W + c_B$.  The operator coefficients $c_{WB}$, $c_T$ and $c^{1221}_{\ell\ell}$ have one-to-one correspondences with $S$, $T$ and $\delta G_F$, given by
\begin{equation}
c_{WB} = \hat{S} = \frac{\alpha}{4 s^2_w} S \,, \hspace{1.5cm}
c_T = \hat{T} = \alpha T\,, \hspace{1.5cm} c^{1221}_{\ell\ell} = -2 \delta G_F \,,
\end{equation}
and the 3-parameter fit can be recovered by simply setting all other operator coefficients to zero.

\begin{table}
\centering
\begin{tabular}{l|l} \hline\hline
$\mathcal{O}_{WB} = \frac{1}{4} gg' H^\dagger \sigma^a H W^a_{\mu\nu} B^{\mu\nu}$  &
$\mathcal{O}_{T} = \frac{1}{2}(H^\dagger \overleftrightarrow{D_\mu} H)^2 $  \\
$\mathcal{O}^{1221}_{\ell\ell} =  (\bar{\ell}_1 \gamma^\mu \ell_2) (\bar{\ell}_2 \gamma_\mu \ell_1)$  &   $\mathcal{O}_{He} = (i H^\dagger \overleftrightarrow{D_\mu} H )(\bar{e} \gamma^\mu e )$  \\ 
$\mathcal{O}_{Hq} = (i  H^\dagger \overleftrightarrow{D_\mu} H )(\bar{q} \gamma^\mu q ) $ &  $\mathcal{O}_{Hu} = (i H^\dagger \overleftrightarrow{D_\mu} H)( \bar{u} \gamma^\mu u )$ \\
$\mathcal{O}'_{Hq} = (i H^\dagger \sigma^a \overleftrightarrow{D_\mu} H)( \bar{q} \sigma^a \gamma^\mu q) $ &  $\mathcal{O}_{Hd} = (i H^\dagger \overleftrightarrow{D_\mu} H)( \bar{d} \gamma^\mu d )$ \\  \hline\hline
\end{tabular}
\caption{A complete operator basis for the $Z$ and $W$ pole observables assuming flavor universality.  We keep the flavor labels on $\mathcal{O}^{1221}_{\ell\ell}$ to distinguish it from the flavor-diagonal ones ($iijj$) which do not contribute to muon decay.    $q$, $l$ are left-handed $SU(2)_L$ doublets, $e$, $u$, $d$ are $SU(2)_L$ singlets.  Note that the convention of $\mathcal{O}_{WB}$ is slightly different from the one in \cite{DeBlas:2019qco}.   
}
\label{tab:op1}
\end{table}
\begin{figure}
    \centering
    \includegraphics[width=0.6\textwidth]{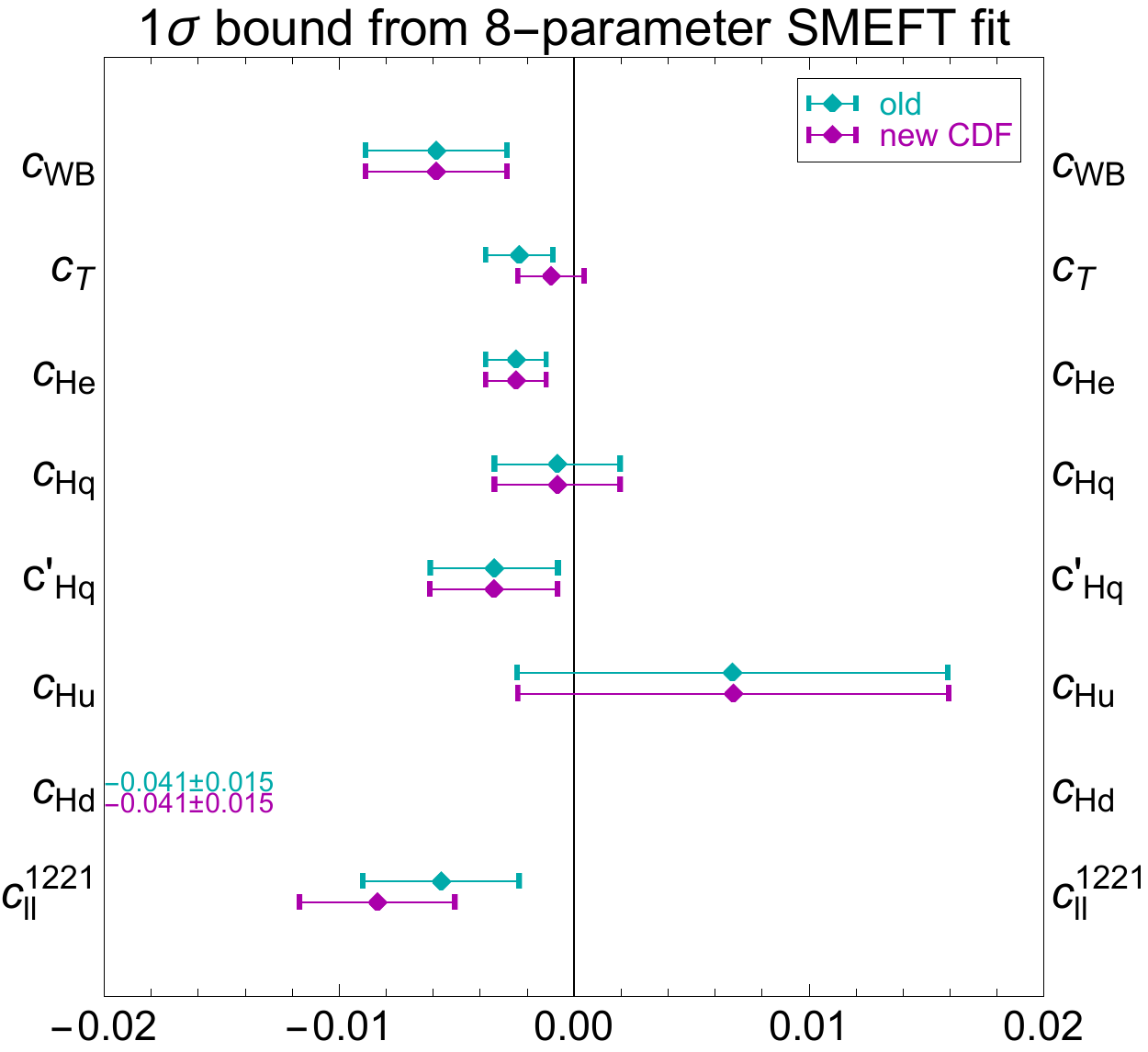}
    \caption{ One-sigma bounds from the 8-parameter SMEFT fit with the operator coefficients listed in \autoref{eq:Lsmeft} and \autoref{tab:op1}.  The ``old'' scenario (cyan) uses the current PDG world-average $m_W$ measurement, while the ``new CDF'' scenario (magenta) uses the new CDF measurement alone for $m_W$.  Note that $c_{Hd}$ is out of the plot range due to the $A^{b}_{\rm FB}$ measurement at LEP.  Its bounds are directly quoted in numbers. 
    }
    \label{fig:para8}
\end{figure}
\begin{table}\small
\centering
\begin{tabular}{|c|c|c|cccccccc|} \hline
& \multicolumn{2}{|c|}{$1\sigma$ bounds (in \%)}  & \multicolumn{8}{|c|}{correlation matrix} \\ \hline
& old & new CDF & $c_{WB}$ & $c_T$ & $c_{He}$ & $c_{Hq}$ & $c'_{Hq}$ & $c_{Hu}$ & $c_{Hd}$ & $c^{1221}_{\ell\ell}$ \\ \hline
$c_{WB}$  &  $-0.59 \pm 0.30$ & $-0.59 \pm 0.30$ & 1 & 0.96 (0.97) & 0.96 & -0.091 & -0.25 & -0.16 & 0.11 & 0.91 \\
$c_T$ &  $-0.23 \pm 0.14$  & $-0.10 \pm 0.14$ &  & 1 & 0.93 & -0.07 & -0.20 & -0.16 & 0.15 & 0.78 (0.80) \\
$c_{He}$ & $-0.25 \pm 0.13$ &  $-0.25 \pm 0.13$  &   &   & 1 & -0.12 & -0.29 & -0.14 & 0.05 & 0.85 \\
$c_{Hq}$ & $-0.07 \pm 0.27$ & $-0.07 \pm 0.27$ &  &   &   & 1 & -0.30 & 0.60 & 0.38 & -0.13 \\
$c'_{Hq}$  & $-0.34 \pm 0.27$ & $-0.34 \pm 0.27$  &   &   &   &   & 1 & -0.69 & 0.58 & -0.33 \\
$c_{Hu}$  & $0.67 \pm 0.92$ & $0.68 \pm 0.92$ &   &   &   &   &   & 1 & -0.07 & -0.11 \\
$c_{Hd}$  & $-4.1 \pm 1.5$ & $-4.1 \pm 1.5$ &   &   &   &   &   &   & 1 & -0.02 \\
$c^{1221}_{\ell\ell}$  & $-0.56 \pm 0.33$ & $-0.84 \pm 0.33$ &   &   &   &   &   &   &   & 1 \\ \hline
\end{tabular}
\caption{The one-sigma bounds from the 8-parameter SMEFT fit (as shown in \autoref{fig:para8}) and the correlation matrix.  The one-sigma bounds are shown in percentages. (A factor of $10^{-2}$ should be applied to all the bounds.) For the correlation matrix, most entries are the same for the two scenarios.  For the different ones, the new CDF numbers are in the parentheses.}
\label{tab:para8}
\end{table}
The $1\sigma$ bounds from the 8-parameter fit is presented in \autoref{fig:para8} and \autoref{tab:para8}.  Again, a comparison is drawn between the ``old'' scenario (with the PDG world-average $m_W$) and the ``new CDF'' one.  
For the ``old'' scenario, our results are in good agreements with the ones in Ref.~\cite{Falkowski:2014tna} for the SILH' basis.  
Caution must be taken in the interpretation of the global fit results, as the introduction of many parameters could result in an over-fitting to the data.  As in Ref.~\cite{Falkowski:2014tna}, here the central values of $c_{WB}$ and $C_T$ becomes negative, and we observe very strong correlations among $c_{WB}$, $c_T$ and $c_{He}$ in the range $0.93$-$0.96$ (see \autoref{tab:para8}), which are also strongly correlated with $c^{1221}_{\ell\ell}$.  The minimum $\chi^2$ in this case is $17.1$, much smaller than the ones of the 3-parameter fit ($28.3$), suggesting a significant shift in the global minimum.  We note also that $c_{Hd}$ exhibits a significant deviation from the SM due to the measurement discrepancy in the $A^b_{\rm FB}$ measurement of about $2.5\sigma$~\cite{ALEPH:2005ab,Baak:2014ora}.\footnote{We also note that the beautiful mirror model~\cite{Choudhury:2001hs} which modifies the $Zb\bar{b}$ couplings to resolve this discrepancy, generally also prefers a positive $T$ parameter in the range $\sim 0.1-0.3$ if the bottom partners are around or above the TeV range~\cite{Gori:2015nqa}.}  However, for the comparison between the ``old'' and ``new CDF'' scenarios, we still observe a pattern similar to the case of the 3-parameter fit, that only significant shifts in $c_T$ and $c^{1221}_{\ell\ell}$ are needed to accommodate the shift in $m_W$.  We also note that the 7$\sigma$ deviation in the CDF $m_W$ measurement is significantly diluted in the 8-parameter fit.  This is expected, as the marginalized bounds become weaker with more parameters.  It is also reflected in the strong correlations among $c_{WB}$, $c_T$, $c_{He}$ and $c^{1221}_{\ell\ell}$, as mentioned above. 
The situation is different for $c_{Hd}$ which is almost in the same direction as $A^b_{\rm FB}$ (or $A_b$)~\cite{Gori:2015nqa}, and basically inherited its $2.5\sigma$ deviation. 





\section{Class of Models}
\label{sec:BSM}

This section explores the compatibility of various representative BSM models that help explain this discrepancy. 

\subsection{$W^\prime$ and $Z^\prime$}
\label{sec:zprime}

We first focus on the simple models with gauge symmetry $SU(2) \times SU(2) \times U(1)$ and discuss the low energy consequences of different symmetry breaking patterns. Those models can be seen as the prototype of some more sophisticated models to explain the origin of mass. 

\subsubsection*{case I: left-handed breaking}
The first case is that a gauge symmetry of $SU(2)_1 \times SU(2)_2 \times U(1)_Y$ ($U(1)_Y$ is the Hypercharge gauge symmetry). And a bifundamental scalar $\Delta$ neutral under Hypercharge can get a VEV to break $SU(2)_1 \times SU(2)_2$ into diagonal subgroup $SU(2)_L$ which is identified as electroweak gauge symmetry,
\bea
\langle \Delta \rangle =\frac{1}{\sqrt{2}}\left(\begin{array}{cc}
    v_\Delta   & 0 \\
  0 &  v_\Delta
\end{array}\right).
\eea
We can find that two gauge boson triplet $W_1^a$ and $W_2^a$ will mix, and mass eigenstate $W^{\prime a}$ and SM gauge triplet $W^a$ can be obtained through the following rotation matrix,
\bea
\left(\begin{array}{c}
W^a \\
W^{\prime a} 
\end{array}\right) =\left(\begin{array}{cc}
    \cos \theta  & \sin \theta  \\
  -\sin \theta &   \cos \theta 
\end{array}\right)\left(\begin{array}{c}
W_1^a \\
W_2^a 
\end{array}\right), 
\eea
where mixing angle is $\sin \theta =g_1/\sqrt{g_1^2 +g_2^2}$. The mass of $W^{\prime a} \equiv (W^{\prime +}, Z^\prime, W^{\prime -})$ triplet is 
\bea
m_{Z^\prime} = m_{W^\prime} = \frac{1}{2}\sqrt{g_1^2 +g_2^2} \,\, v_\Delta. 
\eea

We can introduce a $SU(2)_1$ scalar doublet $H$, which can be identified as SM Higgs doublet, to break electroweak (EW) gauge symmetry. The quantum number under $SU(2)_1 \times SU(2)_2 \times U(1)_Y$ of SM fermions can be assigned as follows 
\bea
L_L \in (2,1,-1/2) \quad L_R \in (1,1,-1), Q_L \in (2,1,+1/6) \quad U_R \in (1,1,+2/3) \quad D_R \in (1,1,-1/3),   
\eea
where $L_{L,R}$ can be identified as SM lepton doublet and singlet, and $Q_L$, $U_R$, and $D_R$ are quark doublet, up quark and down quark (here we neglect the QCD $SU(3)_c$ gauge symmetry). With the above quantum number assignment, we can find that the couplings of $W^\prime$ to the SM currents are universal and can be written as
\bea
\mathcal{L}=-g \tan \theta W^{\prime a}_\mu \left( \bar{L}_L \gamma^\mu T^a L_L  + \bar{Q}_L \gamma^\mu T^a Q_L +( i H^\dagger T^a (D_\mu H) -i (D_\mu H)^\dagger T^a H)  \right),    
\eea 
where $g$ is the EW $SU(2)_L$ gauge coupling $g =g_1 \cos \theta$, $T^a =\frac{1}{2} \sigma^a$, and $\sigma^a$ is Pauli sigma matrices. Then we can integrate out the massive $W^\prime$ triplet at tree level and get the effective Lagrange, 
\bea
\mathcal{L}_{eff} =-\frac{g^2 \tan^2 \theta }{2m_{W^{\prime}}^2}  J^a_\mu J^a_\mu\,, \quad \, J^a_\mu =  \bar{L}_L \gamma^\mu T^a L_L  + \bar{Q}_L \gamma^\mu T^a Q_L +( i H^\dagger T^a (D_\mu H)+ h.c.). 
\eea
We can easily find that there is no correction to $T$ parameter because there is custodial symmetry in this model.
The effective Lagrange can be expanded in terms of Han and Skiba operator bases~\cite{Han:2004az} that contribute to EW precision measurements,
\bea
\mathcal{L}_{eff} = a^\prime ( O_{l^il^j}^t +O_{l^iq^j}^t +  O_{hl^j}^t +O_{h q^j}^t ) + \cdots \,, \quad a^\prime =-\frac{g^2 \tan^2 \theta }{2m_{W^{\prime}}^2 },   
\eea
where the subscript $i,j$ represents the SM fermion generation. 
Using the universal electroweak fit defined in \autoref{sec:EWfit}, our results show that this model tends to make the fit even worse, which is easy to understand from \autoref{eq:fits}, where the additional $W'$ contributions to $\delta G_F$ would reduce the $W$ mass.     



\subsubsection*{case II: right-handed breaking}
In this section, we consider the model with gauge symmetry $SU(2)_L \times SU(2)_R \times U(1)_X$. The $SU(2)_R \times U(1)_X$ is broken into Hypercharge symmetry $U(1)_Y$ at some high scale. And then the electroweak symmetry $SU(2)_L \times U(1)_Y$ is broken by Higgs doublet~\cite{Shu:2011au}. In this model, the Hypercharge is defined as 
\bea
Y=T_R^3+X.
\eea
Here we suppose that $SU(2)_R \times U(1)_X$ is broken by a $SU(2)_R$ triplet with $U(1)_X$ charge $X=1$~\cite{Dobrescu:2015qna,Dobrescu:2015yba},
\bea
\Delta =\frac{1}{\sqrt{2}}\left(\begin{array}{cc}
    \Delta^+/\sqrt{2} & \Delta^{++} \\
  \Delta^{0} &  -\Delta^+/\sqrt{2}
\end{array}\right). 
\eea
The neutral component gets a VEV $\langle \Delta^{0}  \rangle = v_R$, and breaks $SU(2)_R$ guage symmetry. The gauge bosons will get the mass,
\bea
m_{W_R^\pm}^2=\frac{1}{4} g_R^2 v_R^2\,, \quad  m_{Z_R}^2 =\frac{1}{2} (g_R^2 +g_X^2) v_R^2.  
\eea
At some low energy scale, the $SU(2)_L$ Higgs doublet $H$ with $U(1)_X$ charge $X=1/2$ gets the VEV $\langle H \rangle =v_{SM}$ to break EW. Then neutral gauge bosons will mix, and their  mass matrix is given by    

\begin{eqnarray}
\frac{1}{4}
\begin{pmatrix}
 W^{3R}_{\mu} & A^X_{\mu} & W^3_{\mu}
\end{pmatrix}
\begin{pmatrix}
2 g_R^2 v_R^2 & - 2 g_R g_X v_R^2 & 0 \\
- 2 g_R g_X v_R^2 &  g_X^2 (2 v_R^2 + v_{SM}^2) & -g_X g_L v_{SM}^2 \\
0 & -g_X g v_{SM}^2 & 2g^2 v_{SM}^2
\end{pmatrix} 
\begin{pmatrix}
W^{3R}_{\mu} \\ A^X_{\mu} \\ W^3_{\mu}
\end{pmatrix}
 \label{G_MG} \ .
\end{eqnarray}
We suppose the EW scale is much smaller than the $SU(2)_R$ breaking scale, so we have a small parameter $\epsilon = v_{SM}^2/v_R^2 \ll 1$.  
 The mass matrix can be diagonalized by rotation matrix $R$,
\bea
\begin{pmatrix} 
W^{3R}_{\mu} \\ A^X_{\mu} \\ W^3_{\mu}  \cr 
\end{pmatrix} = {\rm \bf R^\dagger }
\begin{pmatrix}  A_\mu \cr  Z_\mu \cr  Z^\prime_\mu \cr 
\end{pmatrix} \,
\eea
where $A$, $Z$, and $Z^\prime$ denote the mass eigenstates. The eigenstate $A$ is the photon, $Z$ is identified with the SM $Z$ boson, while $Z^\prime$ is the heavy neutral gauge boson. The couplings
of this model are related to the electric charge
by
\bea
g_R = \frac{e}{\sin \phi \cos \tw} \, , \quad
g_X = \frac{e}{\cos \phi \cos \tw} \, , \quad
e = g \sin \tw \, ,
\eea
where $\tw$ is the weak mixing angle ($\epsilon \to 0$) and $\phi$ is the additional mixing angle ($\sin \phi \equiv g_X/\sqrt{g_R^2 +g_X^2}$).  
The approximate mass expressions of $Z_\mu $ and $Z_\mu ^\prime$ at linear order of small parameter $\epsilon$ are
\bea
\label{eq:zmass1}
m_Z^2 &=& \frac12 v_{SM}^2 (g_Y^2+g_L^2 ) \left[1 - \epsilon \sin^4 \phi \right] +{\cal O}(\epsilon^2) \ ,\\
m_{Z^\prime}^2 &=& \frac12 v_R^2 (g_R^2+g_X^2 ) \left[1 + \epsilon \sin^4 \phi \right]+ {\cal O}(\epsilon^2) \ ,
\eea
where Hypercharge coupling can be identified as   
\bea
\frac{1}{g_Y^2} = \frac1{g_R^2} + \frac1{g_X^2}\ .
\eea
We can see that ${g_X^2}/{g_Y^2} = 1 + \tan^2 \phi$ and ${g_R^2}/{g_Y^2} = 1 + 1/{\tan^2 \phi}$, so the perturbativity of $g_X$ and $g_R$ requires $0.027 <\tan \phi <36$. The work~\cite{Alguero:2022est} studied a similar model at low energy effective interaction level.  

The quantum number of SM fermions under $SU(2)_L \times SU(2)_R \times U(1)_X$ can be assigned as follows 
\bea \label{eq:SMfield}
L_L \in (2,1,-1/2) \quad L_R \in (1,1,-1), Q_L \in (2,1,+1/6) \quad U_R \in (1,1,+2/3), \quad D_R \in (1,1,-1/3).   
\eea
In this model, we assume that SM fields do not take the $SU(2)_R$ charge, so it is similar to the $2-1-1$ model.   

The mixing matrix {\bf R} has the following approximate form for small $\epsilon$:
\bea
\footnotesize {\rm \bf R} = \left (
\begin{matrix}
\sin \phi \cos \tw &
\cos \phi \cos \tw & \sin \tw \cr \sin \phi \sin \tw
+ \epsilon \frac{\sin^3 \phi \cos^2 \phi}{\sin \tw} & \cos \phi \sin \tw
- \epsilon \frac{\cos \phi \sin^4 \phi}{\sin \tw} & - \cos \tw \cr
- \cos \phi + \epsilon \cos \phi \sin^4 \phi & \sin \phi
+ \epsilon \cos^2 \phi \sin^3 \phi  & - \epsilon \cot \tw \cos \phi \sin^3
\phi \cr 
\end{matrix}
\right )\ ,
\eea
and we can simply derive the SM fermion couplings. 

The couplings of $Z$ and $Z^\prime$ to SM fermion can be written as  
\bea
& & g_{Z_\mu f} = g_X \cos \phi \sin \theta_W Q_X - g \cos \theta_W T_L^3 \nonumber \\
&=& \frac{e}{\sin \tw \cos \tw} \left( \sin^2 \tw Q - T_L^3  -\epsilon \sin^4 \phi Q_X \right) ,
\eea
\bea \label{eq:Zprimef}
& &  g_{Z_\mu^\prime f} = g_X (\sin \phi + \epsilon \cos^2 \phi \sin^3 \phi)Q_X - g (- \epsilon \cot \tw \cos \phi \sin^3 \phi)T_L^3 \nonumber \\
&=& \frac{e}{\cos \tw} \left( \frac{\sin \phi}{\cos \phi} Q_X + \frac{ \epsilon \cos \phi \sin^3 \phi}{\sin^2 \theta_W} ( -T_L^3 +Q \sin^2 \theta_W) \right).
\eea

At the $SU(2)_L$ unbroken phase $\epsilon =0$, the couplings of $Z^\prime_\mu$ to SM fields is only from its mixing with $U(1)_X$, so its coupling is just proportional to $U(1)_X$ charge $g_{Z^\prime f} =e Q_X \tan \phi/\cos \tw  $. After integrating out $Z_\mu^\prime$, the Wilson coefficients of the effective operators will be proportional to $Q_X^2$ of the corresponding currents. The effective Lagrange expanded in terms of Han and Skiba language bases~\cite{Han:2004az} are given by 
\bea \label{eq:Zprimeeff}
\mathcal{L}_{eff} 
&=& \frac{a}{2}\Big(O_h +\frac{1}{2}O_{l^il^j}^s +2O_{e^ie^j}^s +O_{l^ie^j} -\frac{2}{3}O_{l^i u^j}
+\frac{1}{3}O_{l^i d^j} -\frac{1}{3}O_{q^i e^j} -\frac{4}{3}O_{e^i u^j}
+\frac{2}{3}O_{e^i d^j} \,\nonumber \\ 
&-&\frac{1}{2}O_{h l^i}^s +\frac{1}{6} O_{h q^i}^s
+\frac{2}{3} O_{h u^i}^s
-\frac{1}{3} O_{h d^i}^s - O_{h e^i}^s\Big)+\cdots. 
\eea
where $a =-e^2 \tan^2 \phi/(m_{Z^\prime}^2 \cos^2 \tw)$ and the subscript $i,j$ represents the fermion generation. Since electroweak precision measurements do not involve four quark operators, we do not explicitly write them out here. We can see that all the four-fermion operators here are from the $Y$ parameter contribution, where the $Y$ parameter is
\bea
Y = -\frac{a m_W^2}{g_Y^2}.
\eea 
Here $Y$ is defined from the $O'_{2B}$ operator coefficient $-g_Y^2 Y / 2 m_W^2$.

We can find that the custodial symmetry violation effect is
\bea
\frac{a}{2} O_h = - \frac{g_R^2+g_X^2} {2 m_{Z^\prime}^2} \sin^4 \phi |h^\dag D^\mu h)|^2 \,,
\eea 
which is coincident with  $Z$ mass shift $\Delta m_Z^2 = - \epsilon \sin^4 \phi m_Z^2$ in \autoref {eq:zmass1}.  
 The corresponding $T$ parameter from the tree level gauge boson mixing can be obtained,
\bea
T = - \frac{a v^2_{SM}}{2\alpha} = \frac{\epsilon \sin^4 \phi}{\alpha}.  
\eea 


 In this simple case, we find that LHC di-muon resonances search~\cite{CMS:2021ctt} excludes the $95\%$ best-fit parameter space for $m_{Z^\prime} <5 $ TeV, considering recent CDF-II results. However, $Z^\prime$ bounds from LHC searches can be eliminated by introducing some vector-like (VL) $SU(2)_R$  fermion doublets to mix with SM fermions. For example, suppose that a VL lepton $SU(2)_R$ doublet $L^\prime =(\nu^\prime, e^{\prime-})$ with $X$ charge $Q_X =-1/2$ interacts with electron doublet and singlet through the Yukawa couplings of some extra scalars. After the neutral components of these scalars get VEVs (suppose their VEVs  smaller than $v_R$, so the gauge symmetry breaking pattern does not change), $e^{\prime -}$ will mix with left- and right-handed electrons, mixing angles supposed to be $\theta_L$ and $\theta_R$. So SM chiral electrons can also interact with $Z^\prime$ through these mixings, and their coupling $g_{Z\mu^\prime f }$ in \autoref{eq:Zprimef}  will change into
 \bea
 g_{Z^\prime  e_L} =\frac{e}{2\cos \tw}\Big( \cot(\phi) S_{\theta_L}^2 - \tan (\phi)\Big)\,,\;\;\; g_{Z^\prime  e_R} =\frac{e}{2\cos \tw}\Big( \cot(\phi) S_{\theta_R}^2 - \tan (\phi) (1 + C_{\theta_R}^2 )  \Big),
 \eea
 where $g_{Z^\prime  e_{L}}$ ($g_{Z^\prime  e_{R}}$) is the coupling of  (right-) left-handed electron to $Z^\prime$, $S_{\theta_{L,R}} \equiv \sin \theta_{L,R}$ ($C_{\theta_{L,R}} \equiv \cos \theta_{L,R}$), and the overall factor $1/2$ is from $Q_X$ charges of electrons and VL lepton $L^\prime$. The coupling proportional to $S_{\theta_{L,R}}$ is from the mixings between $e_{L,R}$ and $e^{\prime-}$. Notice that we neglect the terms proportional to $\epsilon$ in the above expressions.  We can find that there is a cancellation among the terms in $g_{Z^\prime  e_{L,R}}$, so LHC detection bounds on $Z^\prime$ can be significantly relaxed by tuning the parameters $\phi$ and mixing angle $\theta_{L,R}$. Thus, in this case, there should be plenty of unexcluded parameter space to explain recent CDF anomaly.

\begin{figure}[!h]
    \centering   \includegraphics[width=0.495\textwidth]{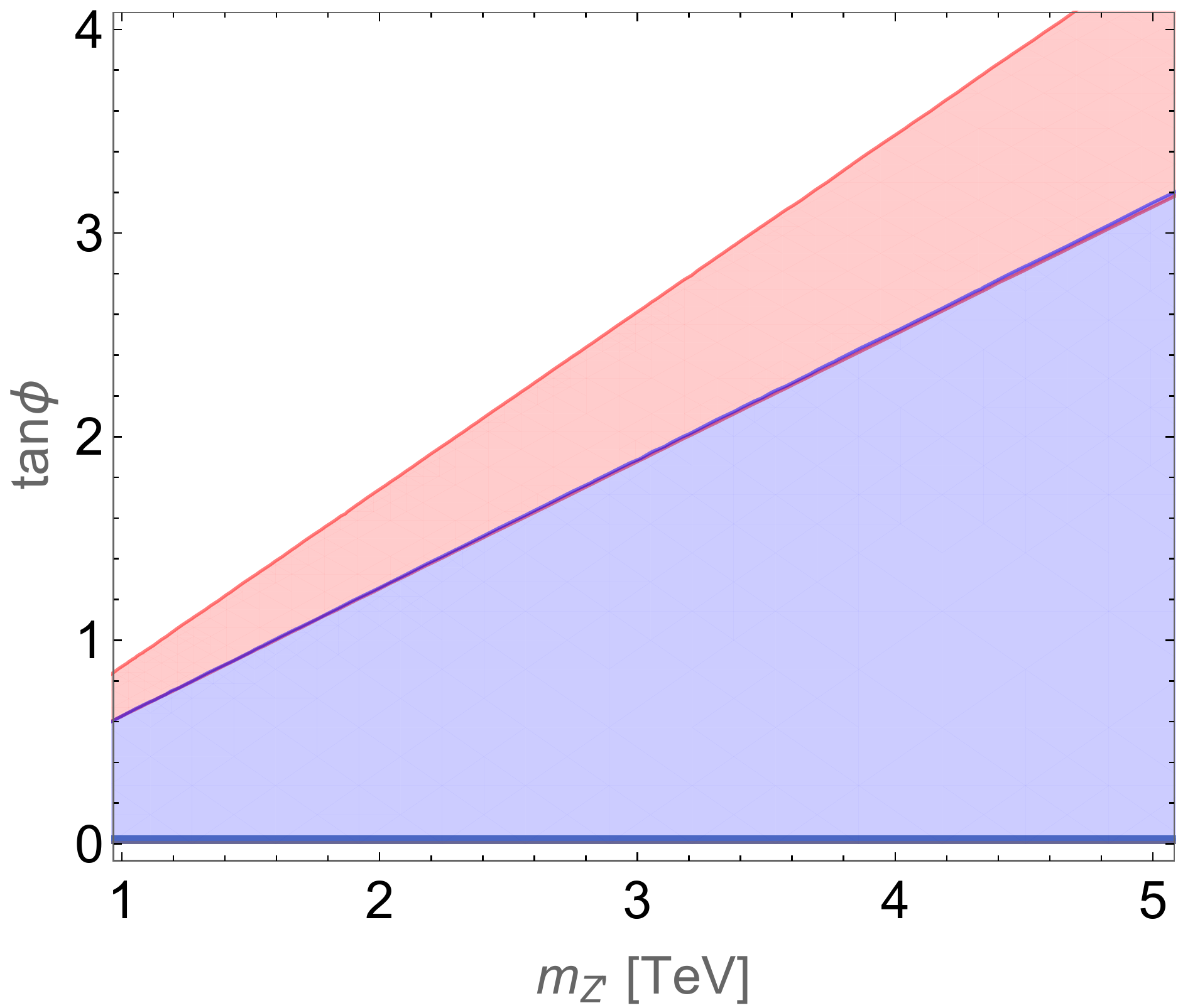}    \includegraphics[width=0.495\textwidth]{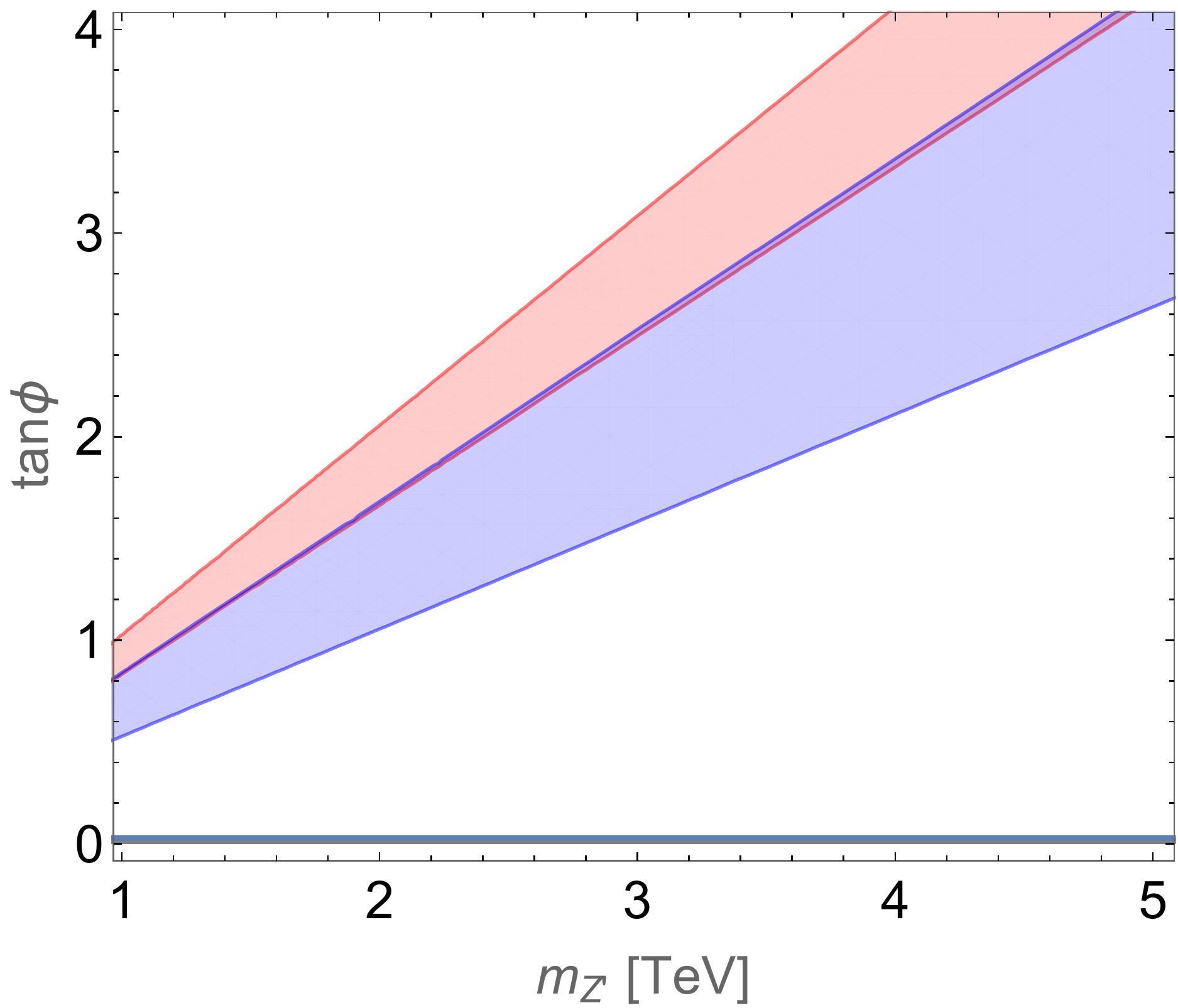}
    \caption{The red and light blue contours correspond to the $95\%$ best-fit band with and without recent CDF-II results. 
    The region below the dark blue line $\tan\phi < 0.027$ is not allowed by perturbation requirement. The left and right panels correspond to extra $S=0$ and $0.1$ cases respectively.
    } 
    \label{fig:Zprime}
\end{figure}

Now we can do the data fit to show the favorite parameter space of recent CDF-II data. As discussed above, in order to get rid of the LHC bounds on $Z^\prime$, we can  assume that the couplings of $Z^\prime$ to SM fermion currents are eliminated by tuning $\phi$ and mixing angles $\theta_{L,R}$ for simplicity, so this model only corrects oblique parameters $S$ and $T$. We calculate the best-fit band (2 $\sigma$ around the local minima of the model) in $\{\tan \phi, m_{Z^\prime } \}$ plane considering the recent CDF-II data, shown in \autoref{fig:Zprime}. Since the loop corrections can contribute to $S$ parameter, we also show the data fit with extra $S=0.1$ contribution in the right panel (the left panel is for no extra $S$ contribution). In the left panel, since total $S=0$, the corresponding $T$ parameter is in the range (0.09, 0.18), which agrees with \autoref{fig:STfit} since the contributions to the global fits from the rest operators are eliminated. While, in the right panel, since the extra positive $S$ contribution is included, the $T$ parameter lower bound is enhanced, data fit preferring large coupling region. 
 

\subsection{Composite Models with Top Partners}

In the composite Higgs model, vector-like composite top partners mix with top quarks to generate top quark mass. In most models, the singlet and doublet top partners always exist. In the following, we simplify top partner models to study their phenomenology generally.    
\subsubsection*{singlet top partner}
\label{sec:toppartner1}
First, we focus on the vector-like quark singlet $T$ with Hypercharge $Y=2/3$ to mix with the top quark. Its general interactions with the top are given by
\bea
\mathcal{L}=-y_t \bar{Q}_L H t_R -y_T \bar{Q}_L H^c T_R -M_T \bar{T}T +h.c. 
\eea
After Higgs get the VEV, we can get the mass of top and singlet top partner,
\bea \label{eq:topmass}
m_t^2 &=& \frac{1}{2}\left(M_T^2 +\lambda_t^2 +\lambda_T^2 -\sqrt{(M_T^2 +\lambda_T^2 +\lambda_t^2)^2 -4M_T^2\lambda_t^2 }\right)\,,\nonumber \\
m_T^2 &=& M_T^2\left(1+\frac{\lambda_T^2}{M_T^2 -m_t^2} \right)\,, \quad \lambda_i =y_i v_{SM} \ ,
\eea 
then we get the EW precision measurement parameters generated from this extra top partner singlet~\cite{Panico:2010is, Marzocca:2012zn}.

After integrate out singlet top partner, we can get the EFT Lagarange in Han and Skiba bases relevant to EW precision measurements,
\bea
a_h&=&-\frac{\alpha}{v_{SM}^2} T\,, \quad  a_{WB}=\frac{\alpha}{8 \sin \tw \cos \tw v_{SM}^2} S\,,\quad \nonumber \\
a_{HQ}^{(1)} &=& \frac{\lambda_{T}^2}{4M_T^2v_{SM}^2} + 
\frac{\lambda_{T}^2 \lambda_t^2}{16 \pi ^2 M_T^2 v_{SM}^4} \left(1 + \log \frac{\lambda_t^2}{M_T^2} \right) 
-\frac{\lambda_{T}^4}{256 \pi ^2 M_T^2 v_{SM}^4} \left( 17 + 14 \log \frac{\lambda_t^2}{M_T^2} \right) 
 \,, \nonumber \\
 a_{HQ}^{(3)} &=& -\frac{\lambda_{T}^2}{4M_T^2 v_{SM}^2} +\frac{\lambda_{T}^4}{256 \pi^2 M_T^2 v_{SM}^4} \left( 9 + 14\log \frac{\lambda_t^2}{M_T^2} \right) \,,    
\eea
where 
\bea
 T = \frac{N_c \lambda_T^2( 2\lambda_t^2 \log (\frac{M_T^2}{\lambda_t^2}) +\lambda_T^2 -2 \lambda_t^2 )}{16 \pi \sin^2 \tw m_W^2 M_T^2},\quad  
S = \frac{N_c \lambda_T^2( 2 \log (\frac{M_T^2}{\lambda_t^2}) -5)}{18 \pi  M_T^2}, 
\eea
$\{a_h, a_{WB}, a_{HQ}^{(1)},a_{HQ}^{(3)}\}$ are the coefficients of bases $\{O_h, O_{WB}, O_{hQ}^s, O_{hQ}^t \}$ in~\cite{Han:2005pr}, where $O_{hQ}^{s,t}$ are the operators involving top doublet (because singlet top partner only interacts with top doublet), and $N_c$ is QCD color number.

\begin{figure}[!h]
    \centering
    \includegraphics[width=0.495\textwidth]{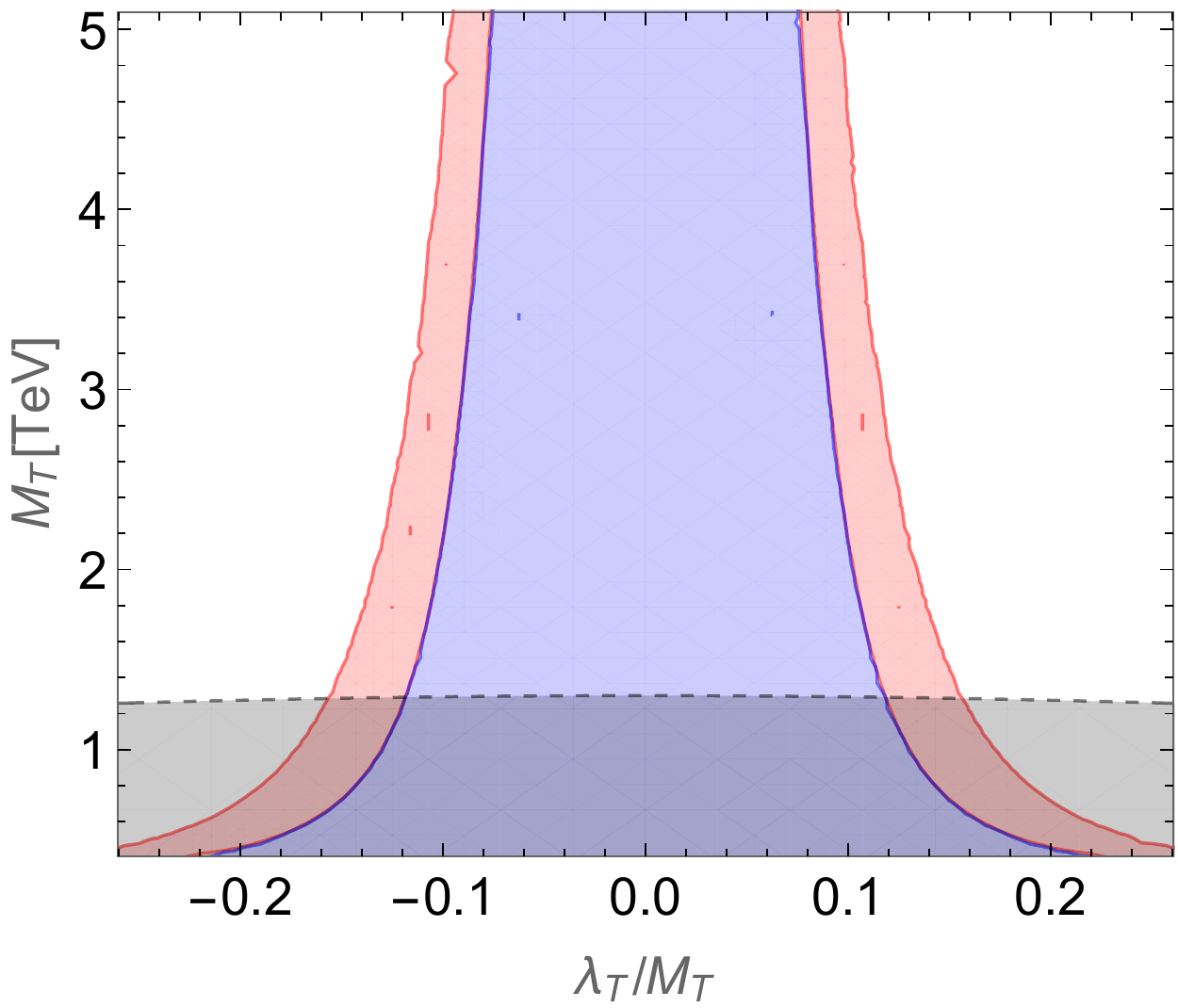}
    \includegraphics[width=0.495\textwidth]{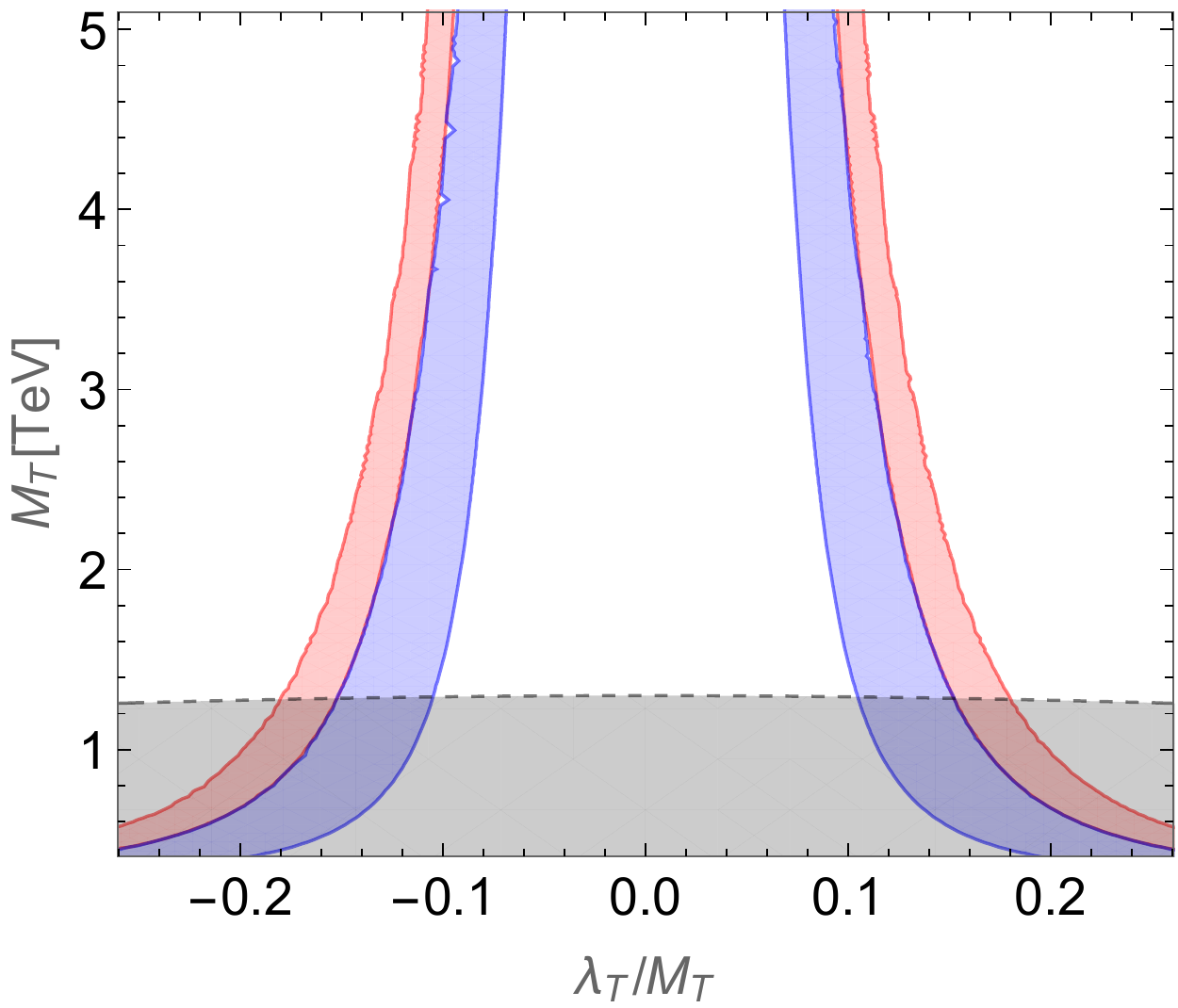}
    \caption{The red and blue contours correspond to the $95\%$ best-fit bands with and without the most recent CDF results. The light dark region is excluded by LHC direct detection   ~\cite{ATLAS:2018ziw}. In the right panel, the contribution of  composite vector partners to $S$ parameter, supposed to be $S=0.1$, is included.  
    } 
    \label{fig:Singlet}
\end{figure}

Similarly, we can obtain the $95 \%$ best-fit bands in $\{M_T, \lambda_{T}/M_T \}$ plane with and without considering the new CDF data  (the red and blue regions in \autoref{fig:Singlet}). In the right panel, we include the extra contribution of the composite vector partners to $S$ parameter in the data fit, supposed to be $S=0.1$.  In the left-panel, since the new CDF data prefers $T>0.1$ (see the solid magenta contour in the left panel in \autoref{fig:STfit}), the red region can not cover the decouple region $\lambda_T/M_T \to 0$. The direct detection of $13$ TeV LHC on singlet top partner excludes the mass region $m_T <1.31$ TeV~\cite{ATLAS:2018ziw}, which corresponds to the black region in \autoref{fig:Singlet}. In the right panel, since the extra $S=0.1$ contribution to EWPT from composite vector meson is introduced, the lower bound of $T$ parameter in \autoref{fig:STfit} is enhanced, which also enhances the lower bounds of coupling $\lambda_T /M_T$ from data fit with and without new CDF data. Since this model does not correct neither the interactions of four light fermions nor gauge interactions of light fermions,  this simple model has a lot of unexcluded best-fit parameter space to explain the anomaly in CDF data. 


\subsubsection*{top partner doublet with $Y= 1/6$}
\label{sec:toppartner2}

\begin{figure}[!h]
    \centering
    \includegraphics[width=0.495\textwidth]{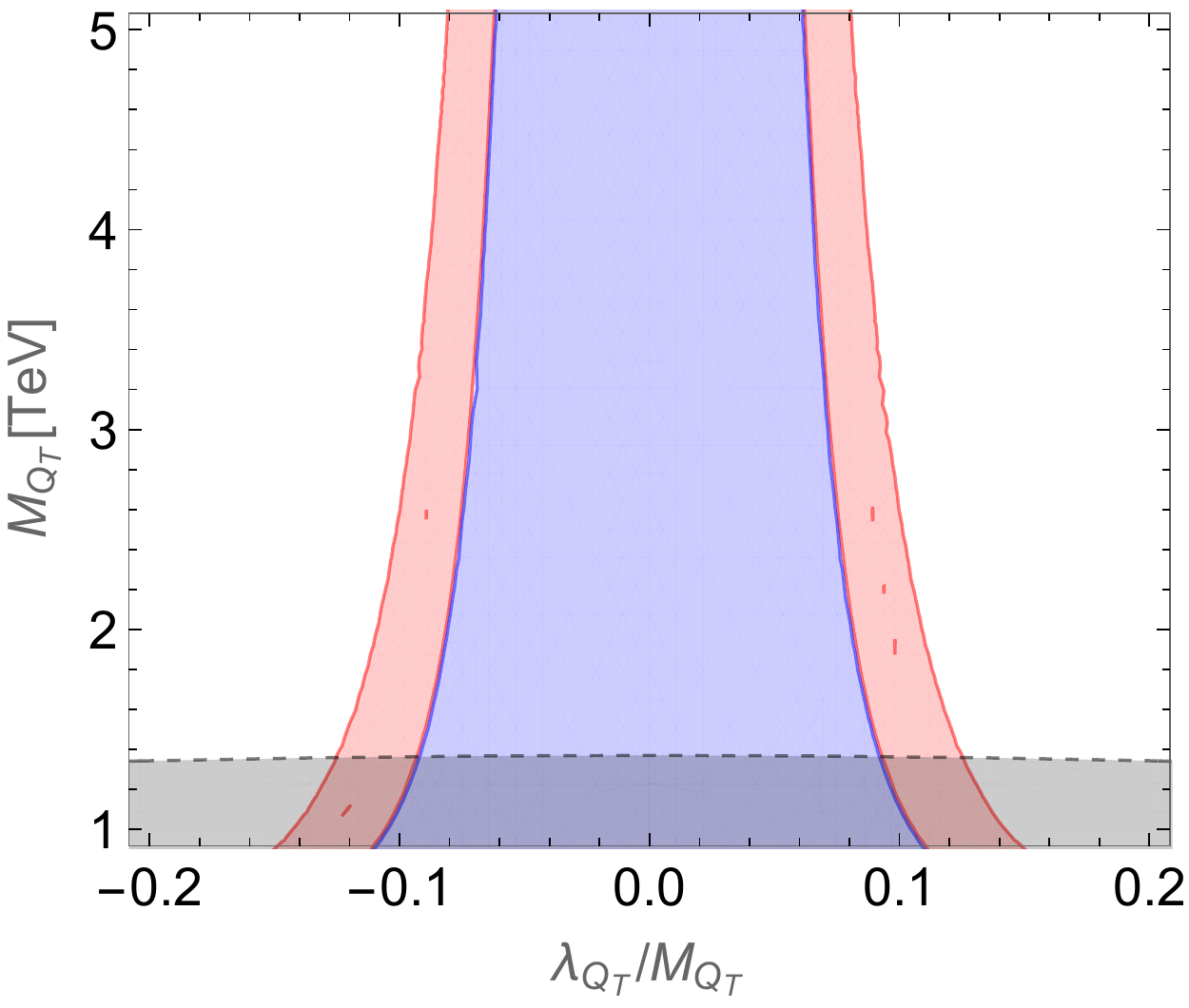}
    \includegraphics[width=0.495\textwidth]{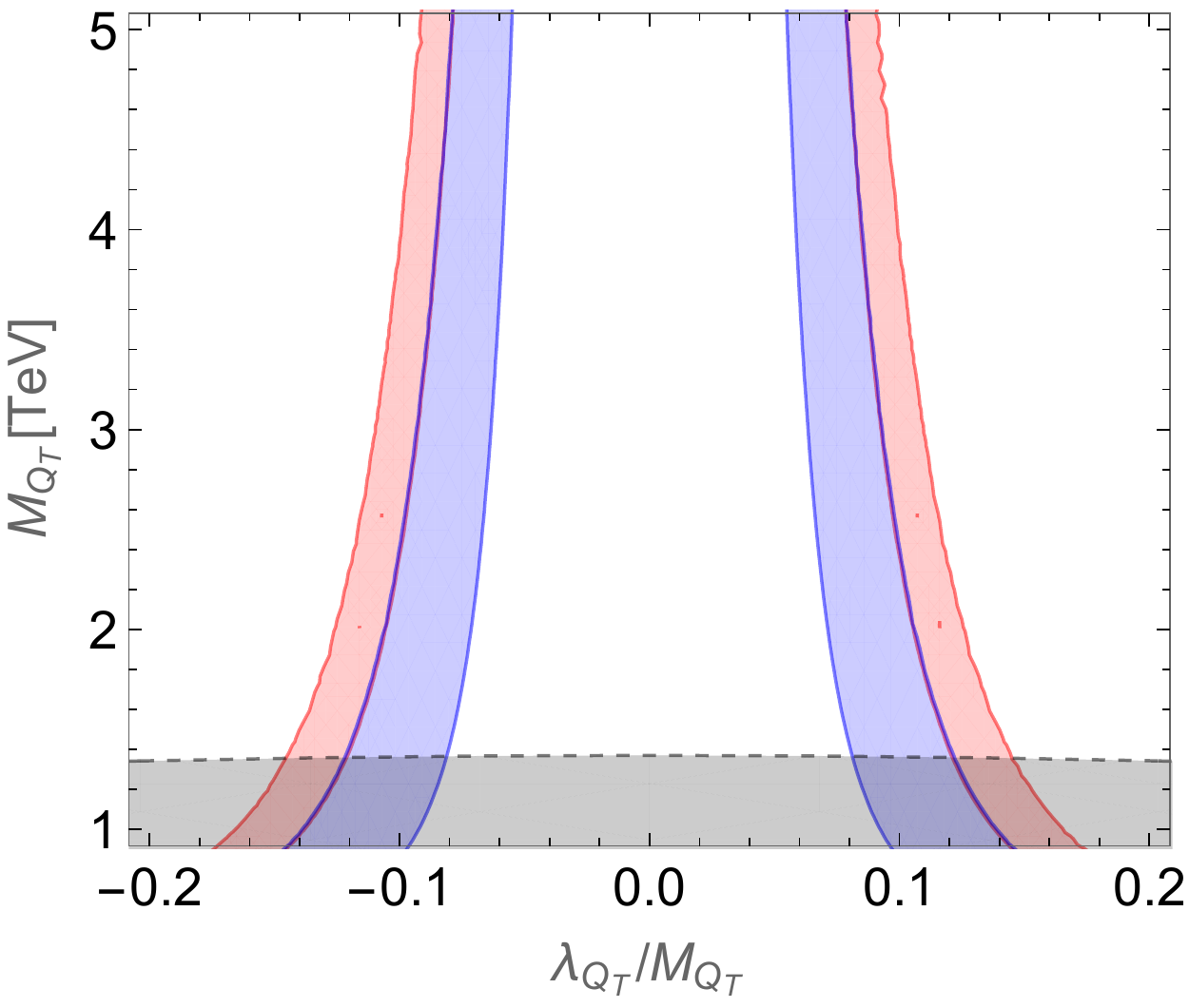}
    \caption{In these two panels, the red and blue contours correspond to the $95\%$ best-fit band with and without the most recent CDF results. The light dark region is excluded by LHC direct detection   ~\cite{ATLAS:2018ziw}. In the right figure, the extra contribution $S=0.1$ from  composite vector partners is included in the data fit.
    } 
    \label{fig:Doublet}
\end{figure}

We can also introduce
vector-like quark doublet $Q_T$ with Hypercharge $Y=1/6$ to mix with top. Its general interactions are given by
\bea
\mathcal{L}=-y_t \bar{Q}_L H t_R -y_{Q_T} \bar{Q}_{TL} H^c t_R -M_{Q_T} \bar{Q}_T Q_T +h.c. 
\eea
The expressions of top and top partner mass are similar to \autoref{eq:topmass},
\bea \label{eq:topmass1}
m_t^2 &=& \frac{1}{2}\left(M_{Q_T}^2 +\lambda_t^2 +\lambda_{Q_T}^2 -\sqrt{(M_{Q_T}^2 +\lambda_{Q_T}^2 +\lambda_t^2)^2 -4M_{Q_T}^2\lambda_t^2 }\right)\,,\nonumber \\
m_{Q_T}^2 &=& M_{Q_T}^2\left(1+\frac{\lambda_{Q_T}^2}{M_{Q_T}^2 -m_t^2} \right)\,.
\eea 
The EW precision measurement parameters from the top partner doublet can be expressed as~\cite{Panico:2010is, Marzocca:2012zn}
\bea
a_h&=&-\frac{\alpha}{v_{SM}^2} T\,, \quad  a_{WB}=\frac{\alpha}{8 \sin \tw \cos \tw v_{SM}^2} S\,,
\nonumber \\
a_{HQ}^{(1)} &=& -\frac{\lambda_t^2 \lambda_{Q_T}^2 }{384 \pi^2 M_{Q_T}^2 v_{SM}^4} \left( 1 + 6\log \frac{\lambda^2_t}{M_{Q_T}^2 v_{SM}^2} \right) \,,\quad  a_{HQ}^{(3)} = -\frac{\lambda_t^2 \lambda_{Q_T}^2 }{96 \pi^2 M_{Q_T}^2 v_{SM}^4}\,,
\eea
where 
\bea
 T &=& \frac{N_c \lambda_{Q_T}^2( 6\lambda_t^2 \log (\frac{M_{Q_T}^2}{\lambda_t^2}) +2\lambda_{Q_T}^2 -9 \lambda_t^2 )}{24 \pi \sin^2 \tw m_W^2 M_{Q_T}^2}\,,\quad  
S = \frac{N_c \lambda_{Q_T}^2( 4 \log (\frac{M_{Q_T}^2}{\lambda_t^2}) -7)}{18 \pi  M_{Q_T}^2}\,.
\eea

Following the same procedures as above, we show the $95 \%$ best-fit bands with and without considering the new CDF data in the red and blue contours in \autoref{fig:Doublet}. In the right panel, the extra contribution $S=0.1$ from composite vector partners is also included. The light black region corresponds to the excluded region $M_{Q_T} < 1.37$ TeV from LHC direct search~\cite{ATLAS:2018ziw}. The physics of this model is almost the same as singlet top partner case, we will not repeat it. Similar to above model, a lot of unexclueded best-fit parameter space in this model can explain new CDF data.    

\subsection{Top Squark}
\label{sec:stop}

Top squark is a representative case for weakly coupled states in the loop that could accommodate the discrepancy. It is instructive and informative to explore the underlying dynamics in this direction. First, we start with a pair of squarks and assume other electroweak states are decoupled. In R-parity conserving supersymmetry, we would need either a neutralino or gravitino to serve as the lightest supersymmetric particles. With a bino-like neutralino consistent with the DM direct detection experiments, we could further check the bino contribution to the low energy effective operators. In the context of R-parity violating supersymmetry, we could have the stop being the LSP can decay promptly or displaced. These R-parity violating operators are typically small enough that they would not affect the EWPOs considered here. Regardless of the top squark lifetime, the LHC put powerful constraints on them. We choose 400~GeV as a minimum requirement on stop mass~\cite{An:2021yqd}. The mass range corresponds to stop having mass splitting with the LSP around 20~GeV, having a shorter lifetime than the reach of the typical disappearing track or displaced vertices search can cover. New searches with soft displaced vertices could help improve the constraints further.

The stop sector with the mass matrix $(\tilde t_L, \tilde t_R)$ 
\begin{eqnarray}
\begin{pmatrix}
m_{\tilde Q_3}^2+m_t^2+D_L & m_t X_t \\
m_t X_t & m_{\tilde u_3}^2+m_t^2+D_R 
\end{pmatrix}
\end{eqnarray}
In the above equation, $X_t$ is related to the SUSY parameters $A_t$, $\mu$, $\tan\beta$ as $X_t= A_t - \mu \cot \beta$.  $D_L$ and $D_R$ are the D-term contributions to top squark masses.

\subsubsection*{Degenerate Top-squark soft masses}

We begin by considering the degenerate stop soft masses, $m_{\tilde Q_3}=m_{U_3}=m_{\tilde t}$. The mass matrix will yield stop mass eigenstates $(\tilde t_1, \tilde t_2)$ with mass eigenvalues of ${m_{\tilde t_1}, m_{\tilde t_2}}$. 

At one-loop order, many operators will be generated. Those most constrained are also relevant for our discussion here, in particular for the $S$-$T$-$\delta G_F$ fit, as~\cite{Henning:2014wua}
\bea
c_T&=& \frac {h_t^4} {64\pi^2 m_{\tilde t}^2} \left[\left(1+\frac 1 2 \frac {g^2 c_{2\beta}} {h_t^2}\right)^2-\frac 1 2 \frac {X_t^2} {m_{\tilde t}^2} \left(1+\frac 1 2 \frac {g^2 c_{2\beta}} {h_t^2}\right) +\frac 1 {10} \frac {X_t^4} {m_{\tilde t}^4} \right]\nonumber\\
c_W&=& \frac {h_t^2} {640\pi^2 m_{\tilde t}^2} \frac {X_t^2}{m_{\tilde t}^2},~~~c_B= \frac {h_t^2} {640\pi^2 m_{\tilde t}^2} \frac {X_t^2}{m_{\tilde t}^2}\nonumber\\
c_{WB}&=& -\frac {h_t^2} {384\pi^2 m_{\tilde t}^2} \left[\left(1+\frac 1 2 \frac {g^2 c_{2\beta}} {h_t^2}\right)-\frac 4 5\frac {X_t^2}{m_{\tilde t}^2}\right]\nonumber\\
c_{2W}&=& \frac {g^2}{320 \pi^2 m_{\tilde t}^2},...
\eea
where $c_{2\beta}\equiv \cos(2\beta)$, $h_t\equiv y_t \sin\beta$.~\footnote{Note that this basis, which is more widely used in the top squark EFT, differs slightly from the basis choice in \autoref{sec:EWfit}. Here we have the $T$ parameter being the same but the $S$ parameter as $\sin^2\!2\theta_W/\alpha~m_Z^2 (4C_{WB}+C_W+C_B)$.}
\footnote{Note that the $c_{2W}$ operator, via equation of motion, can be translated into the four-fermi operator with correct interaction structure that shifts $G_F$. Hence we include it here. However, the size of this operator is very tiny and have negligible impact on the fit.}
Here we omit the details of a few other one-loop generated operators since they have very minor impact on the fit.

\begin{figure}[!h]
    \centering
    \includegraphics[width=0.495\textwidth]{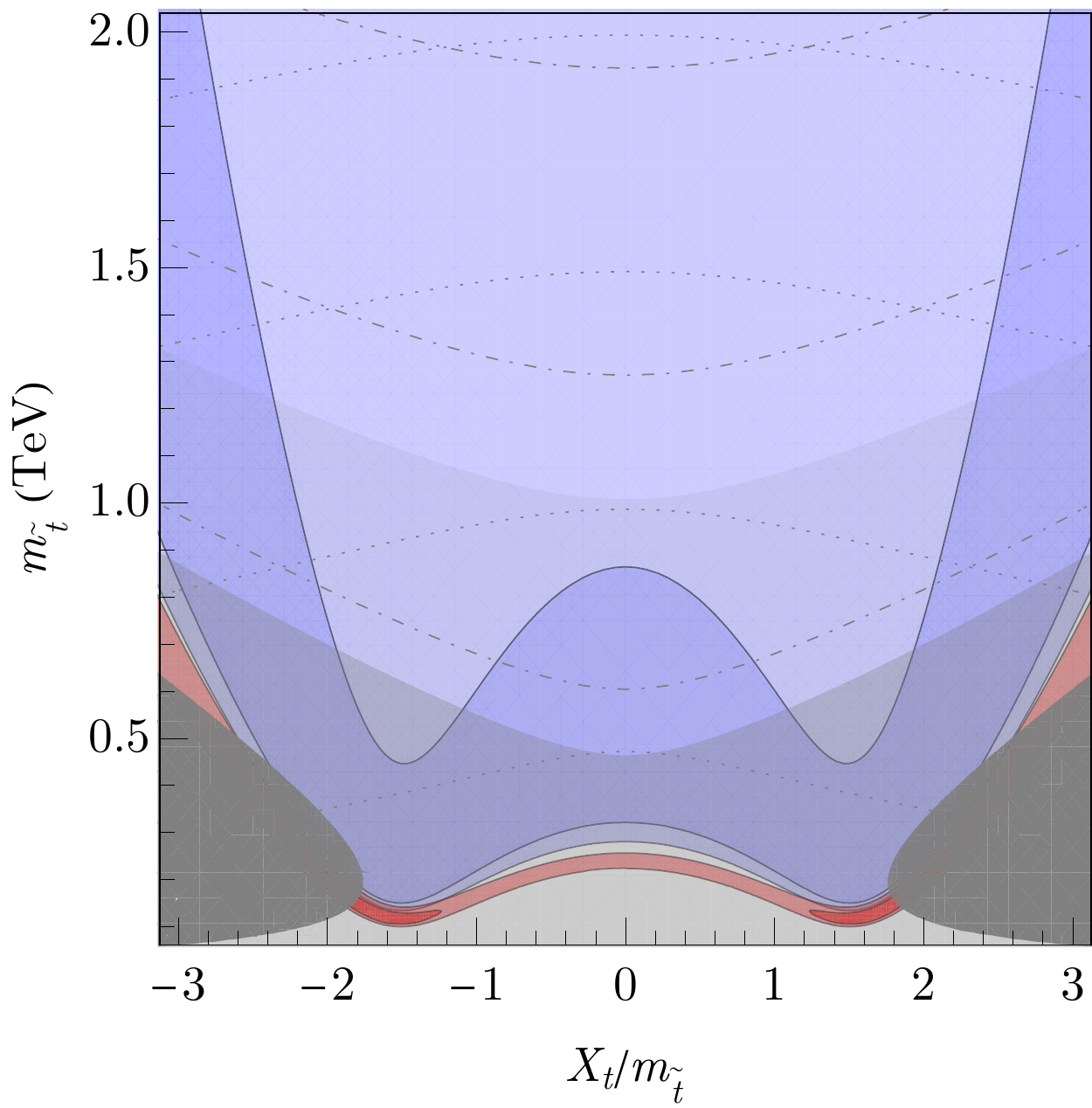}
    \includegraphics[width=0.495\textwidth]{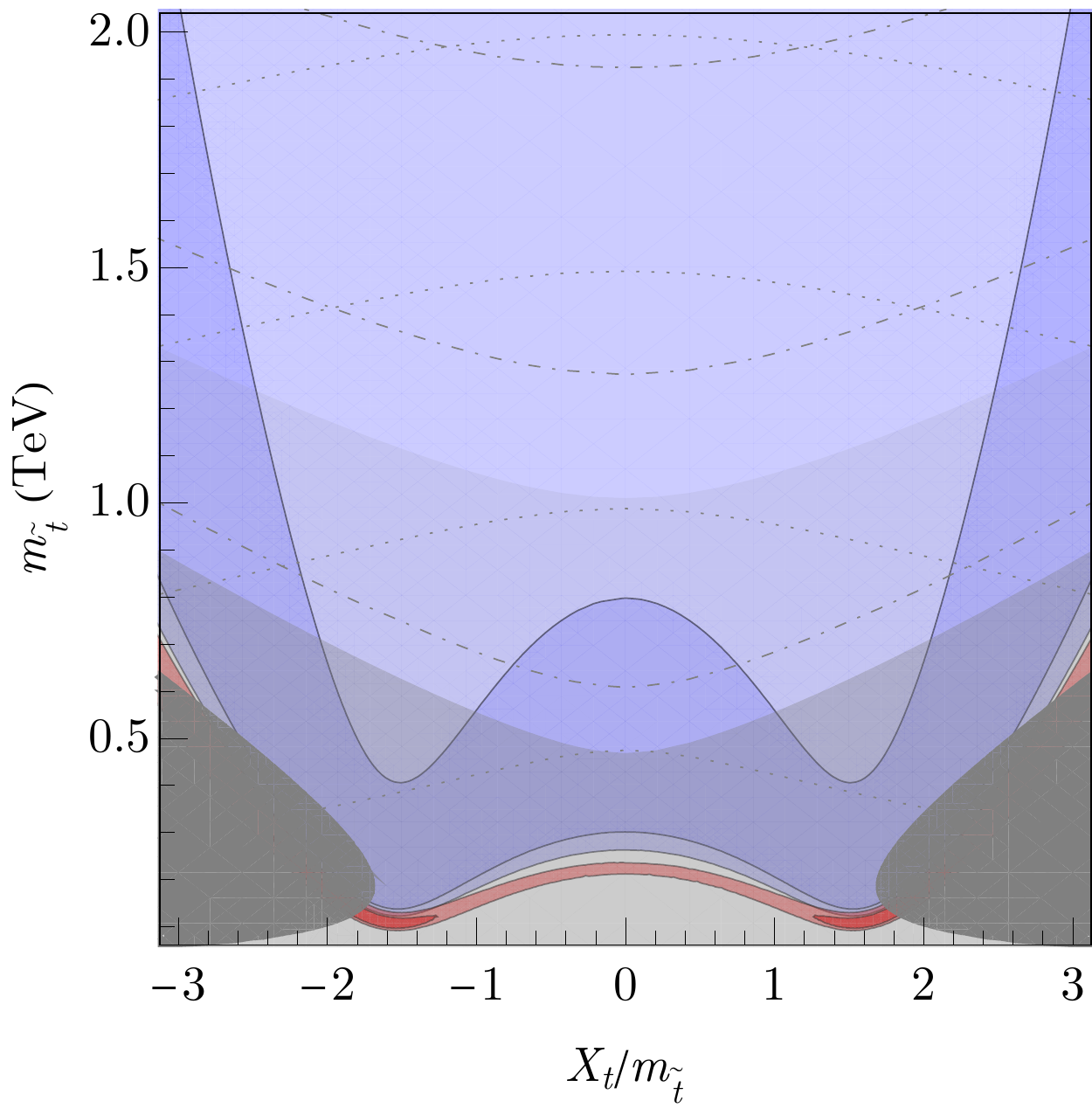}
    \caption{68\% and 95\% fitted parameter range in the top squark parameter space with degenerate stop soft masses, for $\tan\beta=10$ (left panel) and $\tan\beta=3$ (right panel). The recent CDF results require larger $S$ and $T$ parameters to restrain the allowed stop parameter space, as shown in red regions. The pre-CDFII results are shown in blue shaded regions. The gray shaded region is color breaking and hence excluded. The light gray region corresponds to lighter stop mass of 400~GeV and 800~GeV, respectively, representing the current LHC bounds on long-lived stops and prompt stops. The gray dot-dashed and dotted lines represent the top squark mass eigenstates. More details can be found in the text. 
    }
    \label{fig:stop_degenerate}
\end{figure}

With the generated operators from the top squark sector, we can put into $S$-$T$ parameter fit considered in \autoref{sec:EWfit}. We show the 68\% and 95\% best-fit regions with and without the new CDF $m_W$ mass determination in red and blue contours, respectively. The EW fit without CDF results allows for large parameter regions filling the spaces across zero. As anticipated, the recent CDF results require larger S and T parameters to restrain the allowed stop parameter space. The best fit region at 95\% level requires the top squark as light as 200-400 GeV. The required parameter space would be incompatible with other EWPO and Higgs observables. 

The mass scale these solutions point to is so small that one could also worry about the validity of EFT. On the other hand, given that we conclude that top squark with degenerate soft mass terms is insufficient in generating the required shift in oblique parameter, given the current LHC constraints, we can safely discard this possibility.

Still, this provides valuable information to understand the situation with scalar doublets at the loop level. One can consider extending the considerations to multiple specifies, mainly through lepton partners, e.g., light staus~\cite{Agashe:2022uih,Yang:2022gvz}. These can be a promising direction to explore.

\subsubsection*{Non-degenerate Top-squark soft masses}

From the EFT analysis in \autoref{sec:EWfit}, we understand that the ability to generate shifts in $S$ and $T$ parameter directions are crucially important. In particular, from the above discussion, a simple top squark scenario cannot explain the new CDF-II discrepancy between experimental measurements and EW fitted $W$-boson mass. While the top squark sector can already generate custodial symmetry breaking parameters at the one-loop level, one can consider further enhancing these physics effects using the soft mass terms. Non-degenerate top squark soft mass terms will enhance $T$ parameter. We consider a non-degenerate left-handed top squark and right-handed to squark soft mass parameters $m_{Q_3}$ and $m_{U_3}$. 

We define the ratio between the top squark soft mass parameters as
\beq
rr\equiv \frac {m_{U_3}^2} {m_{Q_3}^2},
\eeq
and then we can express the particularly relevant Wilson coefficients as, using standard MSSM parameters already introduced earlier, 
\begin{eqnarray}
c_T&=& \frac {h_t^4} {16\pi^2 m_{\tilde Q_3}^2} \left[\frac 1 4 \left(1+\frac 1 2 \frac {g^2 c_{2\beta}} {h_t^2}\right)^2+2 \frac {X_t^2} {m_{\tilde Q_3}^2} \left(1+\frac 1 2 \frac {g^2 c_{2\beta}} {h_t^2}\right)\left(-\frac {1-5 rr -2 rr^2} {8(1-rr)^3} +\frac {3 rr^2} {4(1-rr)^4} \log(rr)\right)\right. \nonumber\\
&&\left.+ \frac {X_t^4} {m_{\tilde Q_3}^4} \left(\frac {1+10rr+rr^2} {4(1-rr)^4}+\frac {3rr(1+rr)} {2(1-rr)^5} \log(rr)\right)\right]\nonumber\\
c_W&=& \frac {h_t^2} {16\pi^2 m_{\tilde Q_3}^2} \frac {X_t^2}{m_{\tilde Q_3}^2}\left(\frac{1-8rr-17rr^2} {12(1-rr)^4}+\frac {3rr^2+rr^3} {2(1-rr)^5} \log(rr)\right) \nonumber \\
c_B&=& \frac {h_t^2} {16\pi^2 m_{\tilde Q_3}^2} \frac {X_t^2}{m_{\tilde Q_3}^2}\left(\frac{-23-8rr+7rr^2} {12(1-rr)^4}+\frac {-4-12rr+3rr^2+rr^3} {6(1-rr)^5} \log(rr)\right) \nonumber\\
c_{WB}&=& \frac {h_t^2} {16\pi^2 m_{\tilde Q_3}^2} \left[-\frac 1 {24} \left(1+\frac 1 2 \frac {g^2 c_{2\beta}} {h_t^2}\right)+\frac {X_t^2}{m_{\tilde Q_3}^2}\left(\frac {5+33rr-3rr^2+rr^3} {24(1-r)^4}+\frac {2rr + rr^2} {2(1-rr)^5} \log(rr) \right)\right]\nonumber\\
\label{eq:stopwilsonrr}
\end{eqnarray}
and the full list of generated operators can be found here~\cite{Huo:2015nka,Drozd:2015rsp}. In this calculation we decoupled the right-handed bottom squark for simplicity. One can consider adding it and finding better fit to data. Another viewpoint of these loop function can be done in the mass basis, there we can consider the non-degenerate top squark masses generating loop-function variations. 

\begin{figure}
    \centering
    \includegraphics[width=0.7\textwidth]{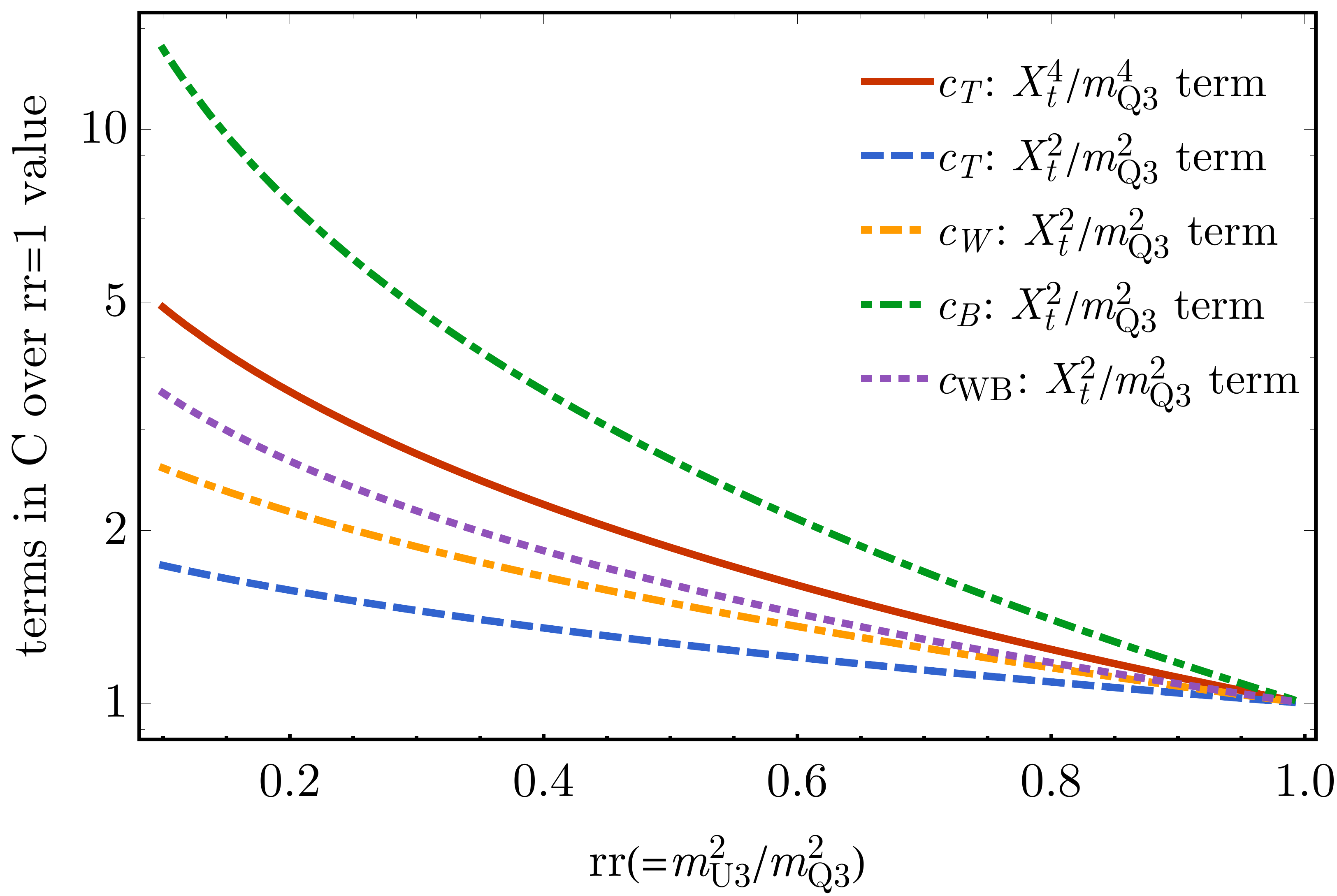}
    \caption{The ratios of various terms in the Wilson coefficients as a function of the soft mass ratios $rr\equiv m_{U_3}^2/m_{Q_3}^2$. The reference values with $rr=1$ for these five terms $c_T (X_t^4/m_{Q_3}^4)$, $c_T (X_t^2/m_{Q_3}^2)$, $c_W (X_t^2/m_{Q_3}^2)$, $c_B (X_t^2/m_{Q_3}^2)$, $c_{WB} (X_t^4/m_{Q_3}^2)$ are 1/40, -1/16, 1/40, 1/40, 1/30, respectively.}
    \label{fig:stopwilson}
\end{figure}

To understand the enhancement, we show in \autoref{fig:stopwilson} the ratios of Wilson coefficients of various terms as a function of $rr$ with respect to the corresponding values of the degenerate case where $rr=1$. For all of $c_T$, $c_W$, $c_B$,$c_{WB}$, the non-degenerate soft-mass $rr$ enters in the L-R mixing term proportional to $X_t^2/m_{Q_3}^2$. For $c_T$, the non-degenerate soft-mass $rr$ enters further in the L-R doubly mixing term proportional to $X_t^2/m_{Q_3}^2$. We can see in this figure that non-degenerate soft mass could provide a factor of a few enhancements in these terms. In particular, the contribution to $c_T$ and $c_B$ are enhanced, which help lift the $S$ and $T$ direction. 

In this \autoref{fig:stopwilson}, the reference values with $rr=1$ for these five terms $c_T (X_t^4/m_{Q_3}^4)$, $c_T (X_t^2/m_{Q_3}^2)$, $c_W (X_t^2/m_{Q_3}^2)$, $c_B (X_t^2/m_{Q_3}^2)$, $c_{WB} (X_t^4/m_{Q_3}^2)$ are 1/40, -1/16, 1/40, 1/40, 1/30, respectively. This ratio is monotonic on $rr$ and become a suppression factor for $rr>1$. Note that this preference implies the light top squark is more preferred to be right-handed, which also makes the top squark parameter space less constrained at the LHC. Another nice feature of these enhancement is for those contributes to $S$ parameter, proportional to $4c_{WB}+c_T+c_B$ all these terms are with a same sign. Further, although the two terms comes with opposite signs for $c_T$, the term and enhancement proportional to $x_T^4/m_{Q_3}^4$ dominants in size in $X_t/m_{Q_3}\gg 1.6$ regions. Hence, the inclusion of non-degenerate top squark soft masses with $rr<1$ will lead to better fit to the new $m_W$ results. 

\begin{figure}[htbp]
    \centering
    \includegraphics[width=0.45\textwidth]{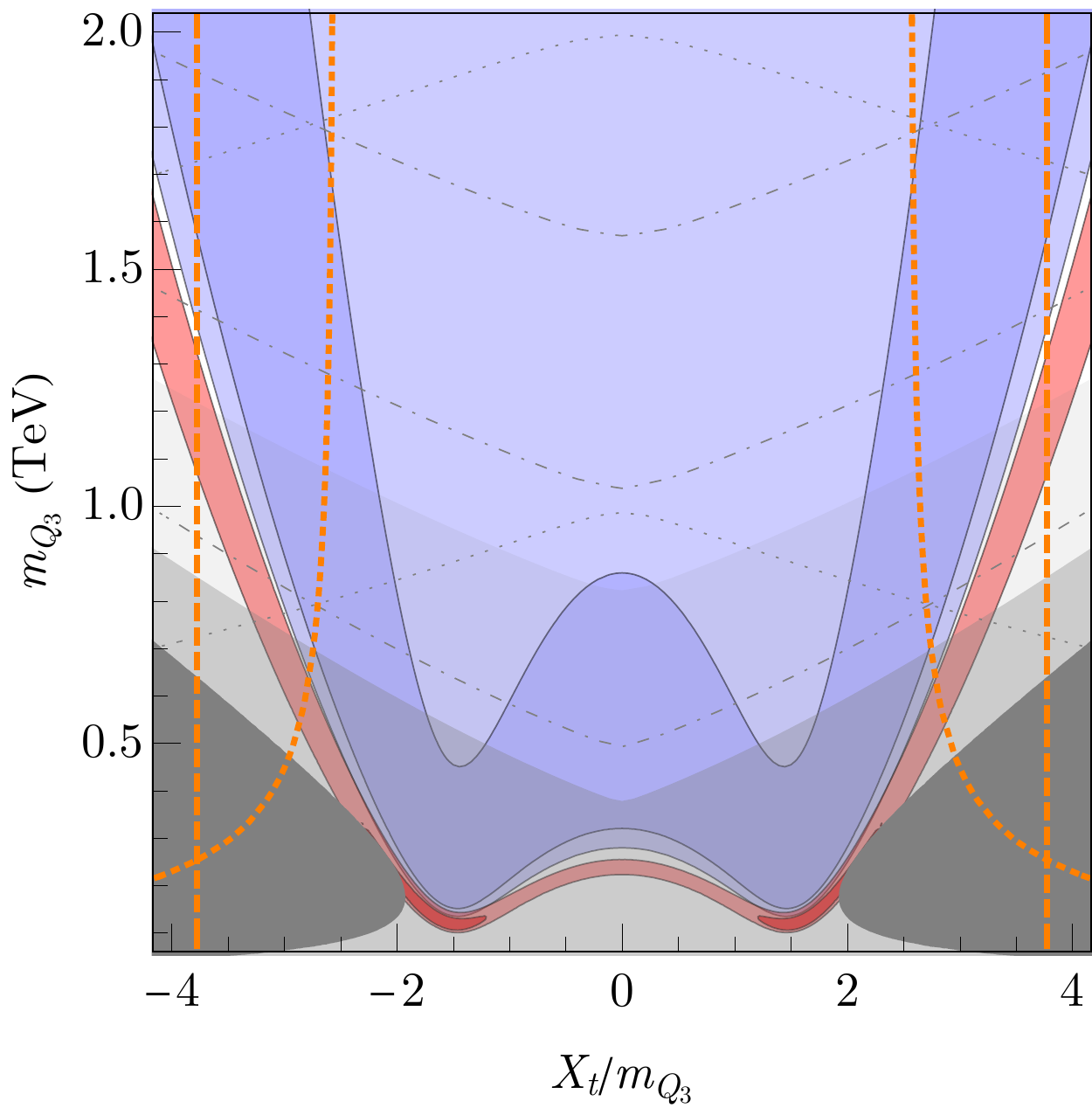}
    \includegraphics[width=0.45\textwidth]{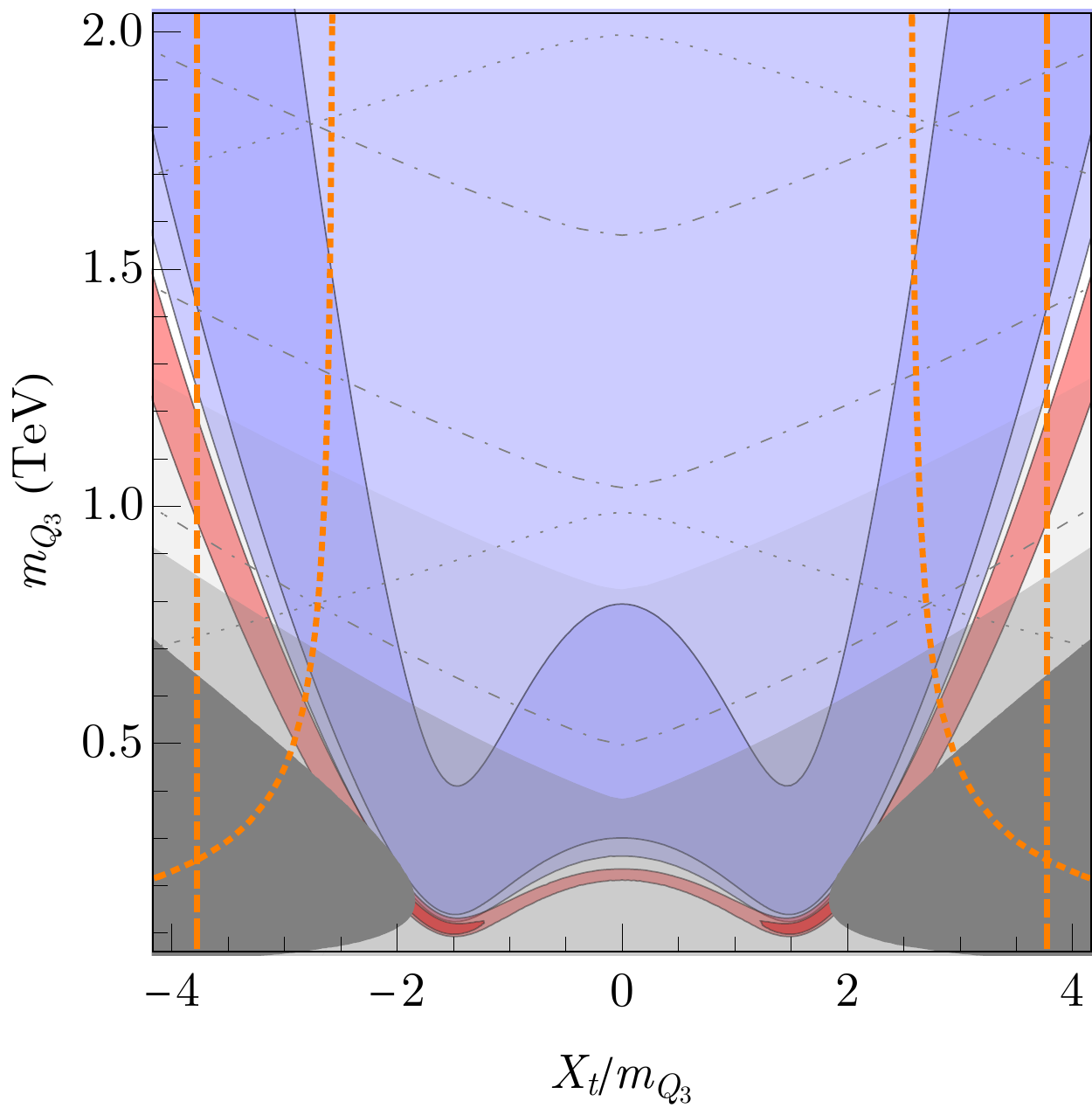}
    \includegraphics[width=0.45\textwidth]{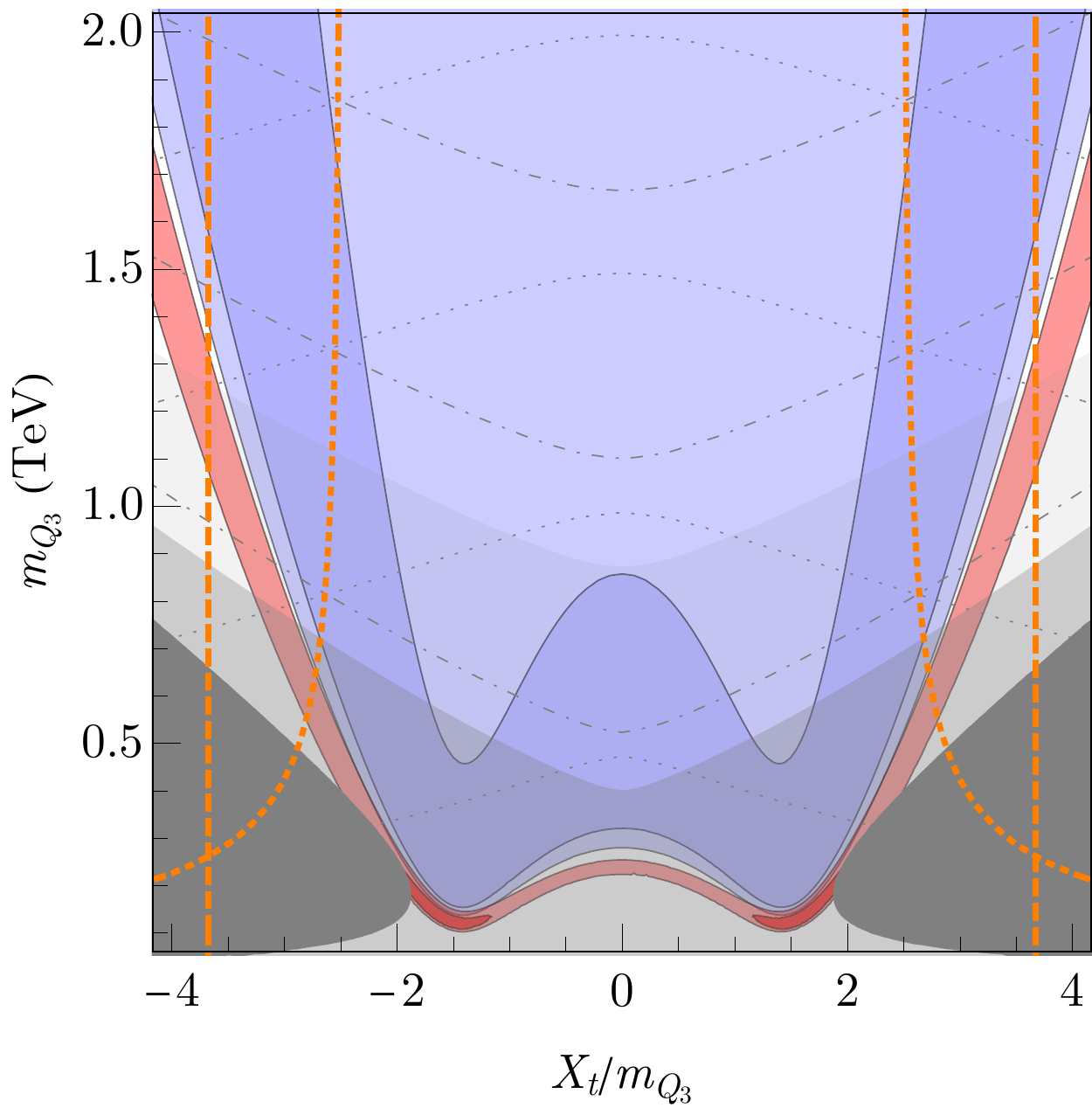}
    \includegraphics[width=0.45\textwidth]{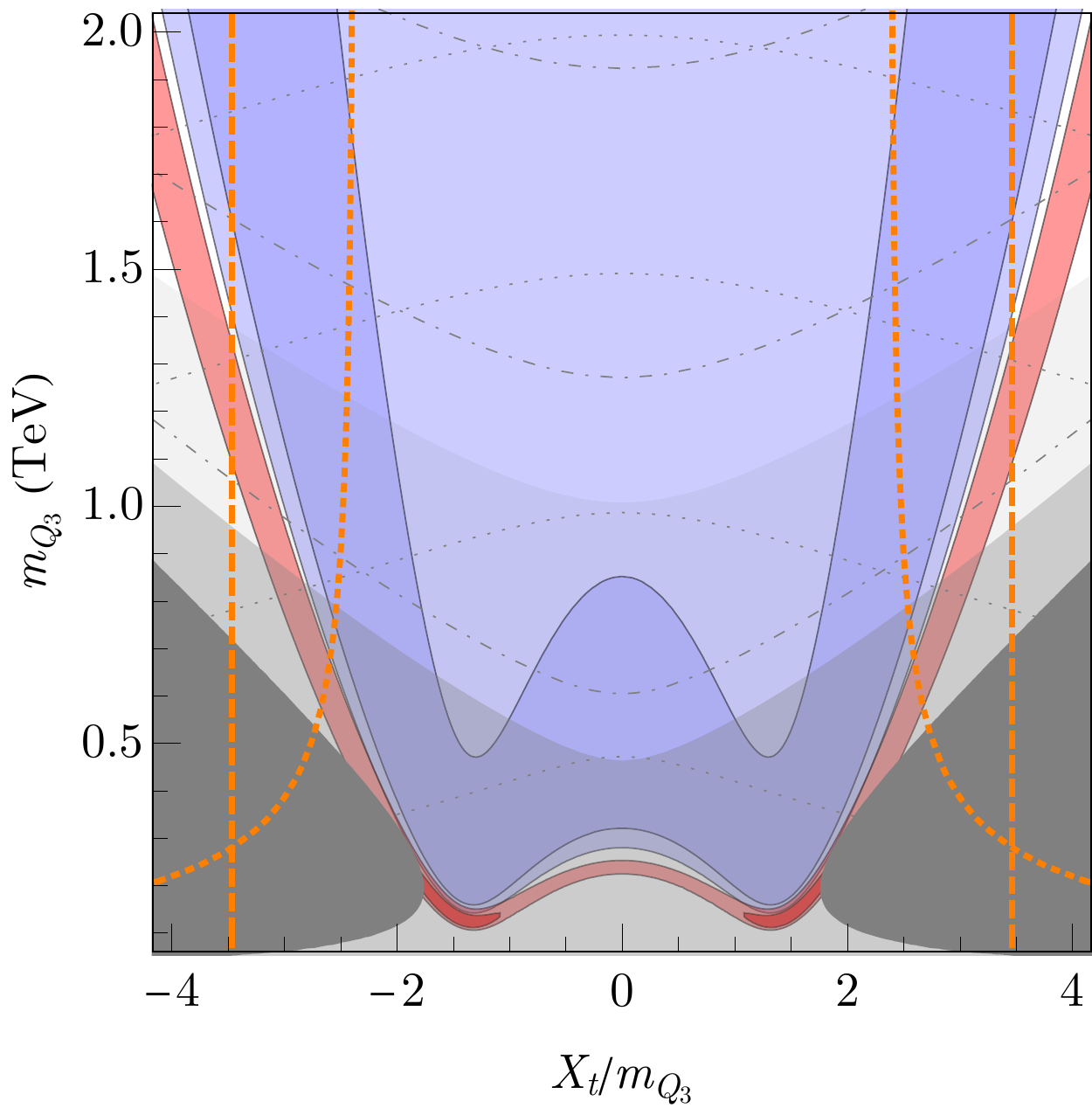}
    \caption{68\% and 95\% fitted parameter range in the top squark parameter space with degenerate stop soft masses. The corresponding top squark parameters are, top-left panel): $rr=0.9$, $\tan\beta=10$; top-right panel): $rr=0.9$, $\tan\beta=3$; bottom-left panel): $rr=0.8$, $\tan\beta=10$; bottom-right panel): $rr=0.6$, $\tan\beta=10$. The blue (red) shaded regions represents the best fit regions with pre-CDFII combined $m_W$ (CDF-II $m_W$ alone).
    The gray shaded region is with tachyonic light stop mass, and the light gray region corresponds to LHC direct search constraints. The dash-dotted and dotted contours label the top squark mass eigenvalues with the separation between them 0.5 TeV.
    The orange lines correspond to two empirical constraints for our metastable electroweak vacuum to tunnel into a color-breaking vacuum, see details in the text. 
    }
    \label{fig:stop_nondegenerate}
\end{figure}

We show the top squark parameter space for various values of $rr$ in \autoref{fig:stop_nondegenerate}. We adopt a similar convention as in \autoref{fig:stop_degenerate}. We emphasize a few important features here. As before, given that CDF-II $m_W$ determination is far away from others and has the best precision, we compare the results with CDF-II $m_W$ alone as the $m_W$ input the results with the $m_W$ before the CDF-II results were reports. The 68\% and 95\% best-fit regions with CDF-II $m_W$ are shown in red shaded regions. Since the new $m_W$ prefers non-zero BSM contributions and is incompatible with the null SM hypothesis, the preferred BSM (red) regions are very restrictive. In contrast, the blue-shaded regions describe the 68\% and 95\% preferred regions with pre-CDFII $m_W$ determination. It covers and allows a complete decoupling direction where $m_{Q_3}$ and goes to infinity and with $X_t$ going to zero.

In \autoref{fig:stop_nondegenerate}, we also use the gray shaded region representing the color-breaking parameter space and hence disallow solutions there. The light gray region corresponds to a light stop mass of 400~GeV and 800~GeV, respectively, indicating the current LHC bounds on long-lived stops and prompt stops. We also provide dash-dotted and dotted contours to label the top squark mass eigenvalues in units of TeV. The lowest lines of each represent 0.5 TeV and 1 TeV, respectively. The difference between contour lines on top squark mass eigenvalues is 0.5 TeV.

We show the results for a few benchmark parameters in the non-degenerate top squark scenarios in this figure. These scenarios are top-left panel): $rr=0.9$, $\tan\beta=10$; top-right panel): $rr=0.9$, $\tan\beta=3$; bottom-left panel): $rr=0.8$, $\tan\beta=10$; bottom-right panel): $rr=0.6$, $\tan\beta=10$. First of all, the results for the non-degenerate case show in \autoref{fig:stop_nondegenerate} are allowed and compatible with current constraints, in contrast to the case degenerate soft-mass results show in \autoref{fig:stop_degenerate}. The newly allowed regions are for $X_t/m_{Q_3}>1$, consistent with the enhancements analysis in \autoref{eq:stopwilsonrr} and \autoref{fig:stopwilson}. 

The results of different $rr$ values are comparatively shown in the top-left panel, bottom-left panel, and the bottom-right panel of \autoref{fig:stop_nondegenerate}. We can see that a lower $rr$ value renders smaller needed $X_t$ values. Such behavior is a natural result of the enhancement associated with small $rr$ values. Hence, small $rr$ values will improve the compatibility with the new CDF-II $m_W$ measurement. On the other hand, a small $rr$ value also implies a lighter right-handed top squark for a fixed value of left-handed top squark soft mass $m_{Q_3}$. There, the color-break vacuum constraints and the LHC direct search constraints are stronger, as can be seen in the gray shaded regions. 
Comparing the top-left panel and top-right panel, we can see the impact of different values of $\tan\beta$. The impact is to drag the band more outwards toward higher $X_t$ for lower $\tan\beta$. Since $h_t$ becomes lower for lower $\tan\beta$, one needs a larger $X_t$ to compensate for the needed operator size. 

Large $X_t$ can generate a new and deeper color-breaking vacuum associated with the scalar direction. We would require our electroweak vacuum to be metastable and have a low zero-temperature tunneling rate longer than the age of our universe. The actual calculation is very involved and depends on many other parameters, and we have empirical and approximated constraints from~\cite{Kusenko:1996jn,Blinov:2013fta}. 
Here in~\autoref{fig:stop_nondegenerate}, we use orange dashed curve to represent empirical constraint from Ref.~\cite{Kusenko:1996jn},
\beq
A_t^2+3\mu^2 < 7.5 (m_{Q_3}^2+m_{U_3}^2).
\eeq
When $A_t^2\gg \mu^2$, one can replace the left-hand side of the above equation with $X_t^2$.
Further, Ref.~\cite{Blinov:2013fta} has derived another approximate constraint, using parameter in this work,
\beq
A_t^2 \lesssim \left(3.4 (1+rr)+0.5 |1-rr|\right)m_{Q_3}^2+60 m_Z^2.
\eeq
Again, when $A_t^2\gg \mu^2$, one can replace the left-hand side of the above equation with $X_t^2$.
We show this constraint in orange dotted lines. We can see such consideration limits us to smaller $X_t/m_{Q_3}$ regions, and a detailed tunneling numerical consideration, with additional parameters in MSSM defined, would help establish the best-fit point in the top squark case.

Overall, we see that non-degenerate top squark soft mass parameters fit the new data much better compared to the degenerate case. Interestingly, the preferred stop parameter direction also yields a significant correction to the Higgs mass via large mixing. Although we do not attempt to fit the observed Higgs mass here, it is well-known that the TeV scale top squark with large mixing could fit it. However, other new physics could also contribute to the Higgs mass. Further, although not yet constraining in the majority of the allowed parameter space, the precision Higgs program can start to play more important roles in the scenarios considered in this study. 






\section{Outlook}
\label{sec:outlook}

The new $W$ mass determination from the CDF-II experiment is remarkable. Together with many other precision measurements, we are stress-testing the concise SM. This intriguing result calls for further explorations in many new directions. Experimentally, new measurements in near and future experiments would help fully establish and converge on the measured values. Theoretically, differential cross sections matching the experimental templates or improved theoretically-clean determination methods shall be explored.  With a joint effort of theory and experiments, the physical meaning of $W$ mass determination will also be 
better clarified, especially when the experimental templates are generated through well-defined and uncertainty-evaluated precision theory calculations. 

Furthermore, as discussed in this work, the tension between the SM and the $m_W$ measurements in the EW fit, if fully established, certainly signals the possibility of new physics. We can explore various BSM scenarios behind such discrepancies. Given the sizable difference in the $W$ mass, the new physics scale needs to be not too far above the TeV scale. Moreover, the new physics could be at the electroweak scale if it generates this discrepancy via loops. Direct new physics searches at the LHC and other experiments will reveal or rule out the new physics model candidates. For instance, one can directly search for new gauge bosons, new top partners, etc., at current and future colliders. The electroweak precision program and the Higgs precision program will also further extract the possible imprints of new physics. 

Finally, the $W$-mass puzzle 
could be easily resolved by a $WW$-threshold scan program at future lepton colliders, such as ILC~\cite{AlexanderAryshev:2022pkx}, CEPC~\cite{CEPCStudyGroup:2018ghi}, FCC-ee~\cite{FCC:2018evy}, and as well C3~\cite{Dasu:2022nux} and muon colliders~\cite{Aime:2022flm}, which will measure $m_W$ to a precision around or even below 1\,MeV. If the discrepancy with the SM is confirmed, such a precise measurement would also point towards a more definite upper bound on the scale of new physics.  High energy future colliders~\cite{FCC:2018vvp,CEPC-SPPCStudyGroup:2015csa} will most likely be able to cover the new physics sources generating such discrepancies. 

\acknowledgements{
The authors would like to thank Majid Ekhterachian and Tony Gherghetta for helpful discussions. We in particular thank Tao Han for helpful discussion in various stages of this work. We also thank Kaustubh Agashe for helpful comments on the manuscript.
J.G. is supported by the National Natural Science Foundation of China (NSFC) under grant No.~12035008. Z.L. is supported in part by the U.S. Department of Energy (DOE) under grant No. DE-SC0022345. T.M. is supported by "Study in Israel" Fellowship for Outstanding Post-Doctoral Researchers from China and India by PBC of CHE and partially supported by grants from the NSF-BSF (No. 2018683), by the ISF (grant No. 482/20) and by the Azrieli foundation.  J.S. is supported by the NSFC under Grants No.
12025507, No. 12150015, No.12047503; and is also supported by the Strategic Priority Research Program and Key
Research Program of Frontier Science of the Chinese Academy of Sciences under Grants No. XDB21010200,
No. XDB23010000, and No. ZDBS-LY-7003 and CAS project for Young Scientists in Basic Research YSBR-006.
}

{\bf Note added:} Recent studies~\cite{Lu:2022bgw,deBlas:2022hdk,Strumia:2022qkt,Athron:2022qpo} also performed EW global fits to the new CDF $m_W$ measurements, which are generally in good agreement with our results.  

\clearpage

\appendix

\bibliographystyle{utphys}
\bibliography{references}

\end{document}